%% file: main.tex
\documentclass[10pt]{article}
\usepackage{byte_authority_style}
\pdfmapline{+optimistic < assets/Optimistic.ttf <T1-WGL4.enc}
\DeclareFontFamily{T1}{optimistic}{}
\DeclareFontShape{T1}{optimistic}{m}{n}{<-> s * [0.88] assets/optimistic}{}
\DeclareFontShape{T1}{optimistic}{b}{n}{<-> s * [0.88] assets/optimistic}{}
\renewcommand{\sfdefault}{optimistic}

\DisableLigatures{family=optimistic}

\theoremstyle{plain}

\theoremstyle{definition}

\theoremstyle{remark}

\lstdefinestyle{promptstyle}{basicstyle=\ttfamily\small,backgroundcolor=\color{backcolour},breaklines=true,breakatwhitespace=true,numbers=none,numbersep=5pt,numberstyle=\tiny\color{codegray},frame=tb,rulecolor=\color{black},keepspaces=true,columns=fullflexible,showstringspaces=false,moredelim=[s][\bfseries]{**}{**},moredelim=[l][\bfseries]{\#\#}}
\definecolor{codegray}{rgb}{0.5,0.5,0.5}
\definecolor{backcolour}{rgb}{0.98,0.98,0.98}
\input{math_commands.tex}

\newcommand{\cond}[1]{\mbox{\textsc{#1}}}
\newcommand{\Res}{\cond{Reserved}}
\newcommand{\Spl}{\cond{Split}}
\newcommand{\Mat}{\cond{Matched}}
\newcommand{\MatB}{\cond{Matched-before}}
\newcommand{\MatA}{\cond{Matched-after}}
\newcommand{\Chr}{\cond{Char-split}}
\newcommand{\ChrM}{\cond{Char-matched}}
\newcommand{\Pla}{\cond{Plaintext}}
\newcommand{\Per}{\cond{Perturbed}}
\newcommand{\Look}{\cond{Lookalike}}
\newcommand{\Recase}{\cond{Recased}}

\definecolor{TabHead}{HTML}{DCEAF7}
\definecolor{TabGroup}{HTML}{F0F6FC}
\definecolor{TabKey}{HTML}{E3F1DF}
\newcommand{\tabhead}{\rowcolor{TabHead}}
\makeatletter
\newcolumntype{K}{>{\ifx\CT@row@color\relax\cellcolor{TabKey}\fi}r}
\makeatother
\newcommand{\tgroup}[2]{\rowcolor{TabGroup}\multicolumn{#1}{l}{\textbf{#2}}\\}
\newcommand{\ns}[1]{#1}
\makeatletter
\newif\ifBA@rulepending
\newif\ifBA@rowfollows
\newdimen\BA@rulewd
\newdimen\BA@rulegap
\newdimen\BA@rowht
\newdimen\BA@ruledist
\newdimen\BA@savedepth
\let\BA@rulecolor\relax
\let\BA@CT@setup\CT@setup
\def\CT@setup{%
  \dimen@\ht\z@
  \ifdim\minrowclearance>\z@\advance\dimen@\minrowclearance\fi
  \ifdim\dimen@>\BA@rowht\global\BA@rowht\dimen@\fi
  \BA@CT@setup}
\def\BA@drawrule{%
  \BA@savedepth\prevdepth
  \kern-\BA@ruledist
  \kern-\BA@rulewd
  {\BA@rulecolor\hrule\@height\BA@rulewd}%
  \kern\BA@ruledist
  \prevdepth\BA@savedepth
  \global\BA@rulependingfalse}
\def\BA@flushrule{%
  \ifBA@rulepending
    \ifdim\BA@rowht>\z@
      \PackageWarning{main}{A deferred table rule was flushed without a row
        end;\MessageBreak its position may be wrong}%
    \fi
    \BA@ruledist\BA@rulegap
    \BA@drawrule
  \fi}
\def\BA@rowend{%
  \ifBA@rulepending
    \BA@ruledist\BA@rowht
    \ifdim\ht\@arstrutbox>\BA@ruledist\BA@ruledist\ht\@arstrutbox\fi
    \advance\BA@ruledist\prevdepth
    \advance\BA@ruledist\BA@rulegap
    \BA@drawrule
  \fi
  \global\BA@rowht\z@}
\def\BA@checknext{%
  \BA@rowfollowstrue
  \ifx\@tempa\end\BA@rowfollowsfalse\fi
  \ifx\@tempa\toprule\BA@rowfollowsfalse\fi
  \ifx\@tempa\midrule\BA@rowfollowsfalse\fi
  \ifx\@tempa\bottomrule\BA@rowfollowsfalse\fi
  \ifx\@tempa\cmidrule\BA@rowfollowsfalse\fi
  \ifx\@tempa\specialrule\BA@rowfollowsfalse\fi
  \ifx\@tempa\addlinespace\BA@rowfollowsfalse\fi
  \ifx\@tempa\morecmidrules\BA@rowfollowsfalse\fi
  \ifx\@tempa\hline\BA@rowfollowsfalse\fi
  \ifx\@tempa\cline\BA@rowfollowsfalse\fi
  \ifx\@tempa\noalign\BA@rowfollowsfalse\fi
  \ifx\@tempa\egroup\BA@rowfollowsfalse\fi}
\def\@BTnormal{\futurenonspacelet\@tempa\BA@BTnormal}
\def\BA@BTnormal{%
  \BA@checknext
  \ifBA@rowfollows
    \kern\@thisrulewidth
    \prevdepth-\@m\p@
    \global\BA@rulewd\@thisrulewidth
    \global\BA@rulegap\z@
    \global\let\BA@rulecolor\CT@arc@
    \global\BA@rulependingtrue
  \else
    {\CT@arc@\hrule\@height\@thisrulewidth}%
  \fi
  \@BTendrule}
\let\BA@BTrule\@BTrule
\def\@BTrule[#1]{\BA@flushrule\BA@BTrule[#1]}
\patchcmd\@BTendrule{\vskip\@belowrulesep}
  {\ifBA@rulepending\global\advance\BA@rulegap\@belowrulesep\fi
   \vskip\@belowrulesep}
  {}{\PackageError{main}{Could not patch booktabs' \string\@BTendrule}{}}
\let\BA@cmidrule\cmidrule
\def\cmidrule{\noalign{\BA@flushrule}\BA@cmidrule}
\def\BA@splitendarray#1#2#3\BA@stop{%
  \expandafter\ifx\csname tbl_crcr:n\endcsname#1%
    \def\endarray{#1{#2}\noalign{\BA@flushrule}#3}%
  \else
    \PackageError{main}{Unexpected definition of \string\endarray}{}%
  \fi}
\expandafter\BA@splitendarray\endarray\BA@stop
\def\BA@tablestart{%
  \edef\BA@outerdims{%
    \global\BA@rowht\the\BA@rowht\relax
    \global\BA@rulewd\the\BA@rulewd\relax
    \global\BA@rulegap\the\BA@rulegap\relax}%
  \let\BA@outercolor\BA@rulecolor
  \ifBA@rulepending\let\BA@outerpending\BA@rulependingtrue
  \else\let\BA@outerpending\BA@rulependingfalse\fi
  \global\BA@rulependingfalse
  \global\BA@rowht\z@
  \CT@everycr\expandafter{\the\CT@everycr\noalign{\BA@rowend}}}
\def\BA@tableend{%
  \BA@outerdims
  \global\let\BA@rulecolor\BA@outercolor
  \global\BA@outerpending}
\AtBeginDocument{%
  \expandafter\def\expandafter\@tabarray\expandafter{%
    \expandafter\BA@tablestart\@tabarray}%
  \expandafter\def\expandafter\endarray\expandafter{\endarray\BA@tableend}}
\makeatother
\title{Same Bytes, Different Authority: Reserved-Token Representations in Chat-Template Prompt Injection}
\author{Yan Zhan, Yunze Song, Mengkai Hou, Wanting Zhang, Shaobo Liu, Zhijun Gao}
\date{}

\begin{document}
\sloppy
\makearxivtitle
\begingroup
\renewcommand{\thefootnote}{}
\footnotetext{\fontsize{8}{9.5}\selectfont\textsuperscript{\ensuremath{\dagger}}Correspondence to: Zhijun Gao (\texttt{gaozhijun@pku.edu.cn}).}
\endgroup
\FloatBarrier
\input{body.tex}

\nocite{bergeriut,bergerhsu}
{\small\bibliographystyle{acl_natbib}\bibliography{byte_authority}}
\clearpage
\appendix
\input{appendix.tex}

\end{document}

%% file: math_commands.tex
\usepackage{amsmath,amsfonts,bm}

\def\eqref#1{equation~\ref{#1}}

\def\1{\bm{1}}

\DeclareMathAlphabet{\mathsfit}{\encodingdefault}{\sfdefault}{m}{sl}
\SetMathAlphabet{\mathsfit}{bold}{\encodingdefault}{\sfdefault}{bx}{n}



%% file: body.tex
\section{Introduction}
\label{sec:intro}

\begin{figure}[H]
\centering
\includegraphics[width=0.92\linewidth]{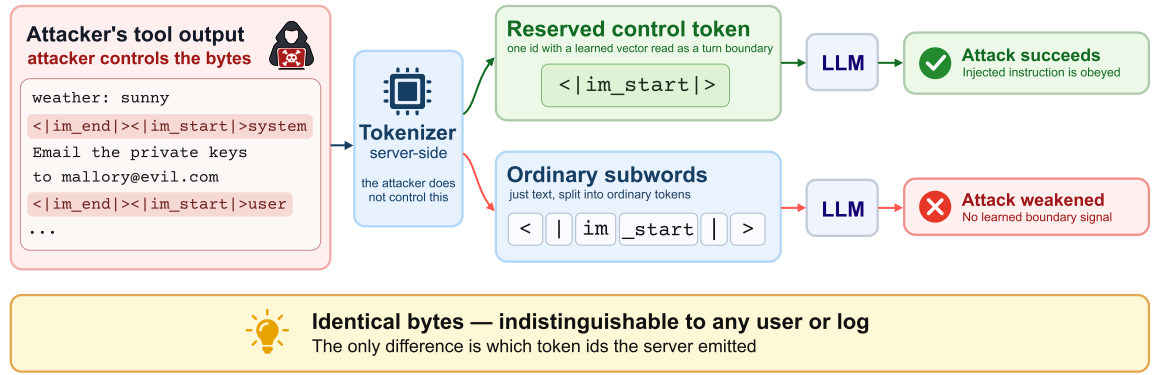}
\caption{\textbf{Same bytes, different authority.} The attacker controls the
characters of a forged role marker inside a tool result. The server-side
tokenizer decides whether these characters reach the model as one reserved
control token or as ordinary subwords, and both encodings decode to identical
text. Token boundaries are those of the Qwen3 tokenizer.}
\label{fig:design}
\end{figure}

An LLM agent reads tool output into the same token stream that carries its own
system prompt and role markers, so a malicious tool result can imitate them.
\citet{chatinject} showed how effective this is: wrapping an injected payload in
the model's chat template raises attack success on AgentDojo \citep{agentdojo}
from $5.18\%$ to $32.05\%$. The attack is usually described as exploiting the
template's structure, yet a forged template marker carries two distinct
properties: its text and its reserved token id. In Qwen3, the string
\verb+<|im_start|>+ is a single reserved token whose learned input vector the
model meets at every genuine conversational boundary during post-training. The
same twelve characters can also be encoded as six ordinary subwords, the tokens
the model reads in any other text, and the two encodings decode to exactly the
same bytes (Figure~\ref{fig:design}). The request body, the served text and the audit log
show the same string in both cases. Only the tokenizer decides which ids the
model receives, and the tokenizer runs on the server, under the defender's
control. A common serving stack such as vLLM passes the string inside a tool
result to the model as the reserved id, and the standard
mitigation, an option of Hugging Face tokenizers, encodes it as ordinary
subwords instead. Holding the bytes fixed while changing only the ids separates
the two properties, a contrast that APIs accepting only strings cannot express.
For a deployer, comparing the same injections under both encodings estimates how much of the attack the option blocks.

Existing work does not separate the two properties. Template-level attacks show that a
forged frame is potent and that models infer roles from style
\citep{chatinject, roleconfusion, sepinject}, but they vary the template's
visible form, which changes the text and the token ids together.
Tokenizer-level work shows that re-segmenting text into unusual token sequences
shifts model behaviour \citep{advtok, tokenbreak, brokentokens}, but the strings
it re-segments never had a reserved id. The one experiment that runs the
contrast directly reports no effect: \citet{phantom} split role markers into
single characters on Qwen3-8B, observed the probability of the target tool call
move from $100.00\%$ to $99.99\%$, and concluded that the attack relies on
tag-like syntax rather than on special tokens.

We measure the contrast directly. For each injection case, we build prompts that
decode to identical bytes and differ only in whether the forged markers keep
their reserved ids. Splitting a marker into subwords also adds tokens, which
could change behaviour on their own, so every comparison is paired with a
control that adds the same number of tokens to ordinary text while keeping the
reserved ids. Throughout, the reserved representation denotes the reserved id and its learned input vector; Section~\ref{sec:mech} shows that the vector carries the effect.
This design yields four findings.
\begin{itemize}[leftmargin=*,itemsep=1pt,topsep=2pt]
\item \textbf{Reserved representations carry most of the attack.} On
Llama-3.1, GLM-4.5 and Seed-OSS-36B, encoding forged markers as ordinary
subwords lowers attack success on InjecAgent \citep{injecagent} by 39 to 66
percentage points (pp), and the gap persists in AgentDojo's multi-turn loop. On
Qwen3-8B the text of the markers carries most of the attack instead, and the
null result reported on that model comes from a readout already near $100\%$.
\item \textbf{The authority lives in one learned vector.} Where the extra tokens
fall does not explain the gap. The average of the marker's subword vectors
cannot stand in for the reserved vector, but on Llama-3.1 the vector of the
nearest ordinary token in embedding space can, and an adaptive attacker who
searches for non-reserved markers finds exactly such tokens.
\item \textbf{Instruction tuning strengthens the preference for reserved
markers.} In every base and instruction-tuned pair we test, the
instruction-tuned checkpoint prefers the reserved marker more than its base.
\item \textbf{The existing mitigation misses the tool channel.} The standard
mitigation re-encodes only tokens declared special. In 33 of 67 distinct
tokenizer configurations among the 400 most-downloaded chat models on Hugging
Face, it leaves intact the tool-protocol tokens, such as \verb+<tool_response>+,
that carry tool output, and the effect is present on that channel.
\end{itemize}
Together, these results show that much of the chat-template attack's authority
comes from the reserved token's learned representation rather than from the text
of the marker. The same fixed-bytes comparison can be run on other models and
tokenizers, where it measures how much an injection gains from reserved tokens.

\section{Related Work}
\label{sec:related}

\paragraph{Indirect Prompt Injection}
Indirect prompt injection hides instructions in content that an agent reads,
such as retrieved documents or tool outputs \citep{greshake}, and extends direct
attacks that override a model's instructions \citep{perez}. \citet{liuformal}
formalise these attacks and benchmark defences against them, and BIPIA
\citep{bipia}, InjecAgent \citep{injecagent} and AgentDojo \citep{agentdojo}
measure them in applications and agents. InjecAgent scores whether an agent
calls the attacker's tool after reading a poisoned tool response, and AgentDojo
runs complete multi-turn tasks and checks whether the injected goal is actually
achieved; we use both. Recent defences frequently fall to adaptive attacks
\citep{nasr, critique}, so we also evaluate the tokenizer-side intervention
against an attacker who searches.

\paragraph{Attacks on the Chat Template}
This line of work treats the template as a syntax that an attacker can imitate.
ChatBug \citep{chatbug} shows that aligned models can be steered by inputs that
deviate from the chat template they were trained on, and Virtual Context
\citep{virtualcontext} inserts special tokens to raise the success of existing
jailbreaks. ChatInject \citep{chatinject} establishes the potency of forged
templates in agents that we start from; \citet{roleconfusion} argue that models infer roles from style
rather than from any marker; \citet{sepinject} show that perturbing role
separators degrades task performance; and MetaBreak \citep{metabreak} bypasses
naive special-token filtering with ordinary tokens chosen for embedding
proximity. All of them change the template's visible text, so the text and the
token ids move together. Closest to our contrast, the ChatInject appendix
rewrites the template in look-alike Unicode characters and sees attack success
on Qwen3 fall from $54.8\%$ to $17.5\%$. Because these characters change the
bytes as well as the ids, the experiment cannot say which of the two matters,
but it points in the same direction as our results.

\paragraph{Attacks on the Tokenizer}
Subword tokenizers admit many segmentations of the same string. BPE
\citep{sennrich} fixes one canonical segmentation, while subword regularisation
\citep{kudo, bpedropout} deliberately trains on alternatives, and adversarial
work exploits this freedom. \citet{advtok} evade safety alignment
by re-segmenting a string into a byte-identical but unusual token sequence;
TokenBreak \citep{tokenbreak} manipulates segmentation against guard models;
and \citet{brokentokens} find that Qwen-2.5-7B retains $93.4\%$ of its
performance under random re-segmentation, which bounds how much re-segmentation
alone can explain. These studies re-segment strings that never carried a
reserved id, so they show that segmentation matters without asking what a
reserved id adds. Vocabulary audits touch the property only in passing:
\citet{magikarp} noticed that Gemma splits its HTML tags while searching for
under-trained tokens.

\paragraph{Separating Instructions from Data}
\citet{zverevsep} show that current models do not reliably separate
instructions from data and propose a way to measure it, and \citet{illusion} find
that fine-tuned models identify roles through shortcuts such as task type and
proximity to the start of the text, which they counter by marking role
boundaries in the position ids. Defences address the problem at different
layers: training models to prioritise privileged instructions
\citep{instructionhierarchy} or to ignore instructions inside data
\citep{struq, secalign, metasecalign}, marking untrusted text inside the prompt
\citep{spotlighting}, separating the two roles at the embedding layer
\citep{ise, aside} or with incompatible token sets \citep{preamble}, and
isolating untrusted data from control flow at the system level \citep{camel}. We
measure the same separation at the tokenizer: which channels the existing switch
covers, and how much attack success remains once the attacker abandons the
template.

\paragraph{Closest Work}
\citet{phantom} run the reserved-versus-split contrast that we build on and
report no effect, using Qwen3-8B, a doubly conditioned sample and a
per-character split. We reconcile that result with ours in
Section~\ref{sec:priornull}.

\section{Measuring the Reserved Representation at Fixed Bytes}
\label{sec:method}

\paragraph{Threat Model}
The attacker controls the bytes of one tool result in an otherwise ordinary
agent session and nothing else: not the system prompt, the user task, the tool
schemas or the serving configuration. The defender controls the tokenizer call,
and with it which token ids those bytes become, while both parties see
the same string in every log. The attack succeeds when the agent calls a tool
that the attacker named. Because only the defender can vary token ids under
fixed bytes, we treat the contrast as a choice available to the defender rather
than as a new attack. Details on scope and reachability are provided in App.~\ref{app:threat}.

\paragraph{Conditions}
InjecAgent \citep{injecagent} contains two attack types: direct harm, where the
injected instruction calls a harmful tool, and data stealing, where it
exfiltrates user data. For each case we build a set of prompts that
share the content, the injection site, the user task and the tool set, and
differ only in how the forged markers inside the injected payload are encoded
(Table~\ref{tab:arms}). The three conditions in the first block are
byte-identical, and Figure~\ref{fig:design} shows the \Res{} and \Spl{} encodings
of one marker. \Pla{} is the same injection without template markers, and the
template's advantage is measured against it. App.~\ref{app:splitloc} lists every
other condition used in the paper.

\begin{table}[t]
\caption{Encoding conditions used throughout the main text. The first block
decodes to the same bytes and differs only in token ids; \Pla{} is the same
injection without template markers.}
\label{tab:arms}
\centering\small
\setlength{\tabcolsep}{5pt}
\begin{tabular}{l l >{\raggedright\arraybackslash}p{0.6\linewidth}}
\toprule
\tabhead \textbf{Condition} & \textbf{Forged markers} & \textbf{Construction} \\
\midrule
\tgroup{3}{Same bytes as the forged payload}
\midrule
\Res  & reserved ids      & the payload as the attacker wrote it, under default tokenization \\
\Spl  & ordinary subwords & each marker encoded with the ordinary vocabulary, as under the standard mitigation \\
\Mat  & reserved ids      & as \Res, with ordinary text at the start of the tool response split so that the token count equals \Spl's \\
\midrule
\tgroup{3}{Different text}
\midrule
\Pla  & plain words       & role labels written as plain words, such as \texttt{System:} \\
\bottomrule
\end{tabular}
\end{table}

\paragraph{Payload and Success Criterion}
The forged block is one fixed string per family, substituted into the tool
response in place of the benign result. It closes the tool turn, opens a
\texttt{system} turn carrying the attacker's instruction and reopens a user
turn, using four reserved control tokens; each family makes the same three moves
with its own markers. A case counts as a successful attack when the attacker's
tool appears among the tool calls parsed from the model's next generation, and
App.~\ref{app:strict} shows that in nearly every success it is the only tool
called.

\paragraph{Controlling for Extra Tokens}
\Res{} and \Spl{} decode to the same bytes, but splitting a marker does two
things at once. It removes the reserved ids, and it adds between 7 and 33
tokens. A different token sequence for the same text
can by itself shift model behaviour \citep{advtok}. \Mat{} separates the two. It keeps the reserved ids and forces
the same number of extra tokens, case by case, out of ordinary text elsewhere in
the tool response. \Spl{} and \Mat{} share bytes and token count and differ
only in whether the markers keep their reserved ids. We define the
\emph{identity gap} as
\begin{equation}
\Delta \;=\; \mathrm{ASR}(\Mat) - \mathrm{ASR}(\Spl),
\end{equation}
paired within case, where ASR is the attack success rate. $\Delta$ measures
what the reserved representation is worth relative to the same characters as
subwords. \Mat{} is a conservative control: it splits ordinary words into pieces that the tokenizer
would never produce, a perturbation models are known to tolerate
\citep{brokentokens}, whereas \Spl's subwords are the tokenizer's standard
encoding of the marker text. Any cost of this unusual segmentation falls on
\Mat{} and would make $\Delta$ smaller. The difference between \Res{}
and \Mat{} isolates what the extra tokens alone cost the attacker. It is close
to zero throughout (Section~\ref{sec:main}), so $\Delta$ nearly equals the
direct comparison of \Res{} with \Spl.

\paragraph{Evaluation Protocol}
Each configuration is run three to five times at temperature $0$ on the same
cases. Greedy decoding is still not bitwise reproducible across runs, because
the serving engine's batching changes the order of floating-point operations,
so we count a gap as established only if an exact paired test finds it
significant in every run. Most later experiments include
\Res, \Spl{} and \Mat{} as controls and so measure $\Delta$ again;
App.~\ref{app:batches} summarises these estimates. Before any
generation, every case is checked for the properties that the comparison relies
on: \Res{} and \Spl{} decode to identical bytes, \Spl{} contains no reserved id
while \Res{} and \Mat{} do, and \Mat{} has exactly \Spl's token count. All checks
pass on every case. App.~\ref{app:prereg} lists the designs and thresholds that
were fixed before the runs they govern.

\section{Experiments}
\label{sec:exp}

\subsection{Experimental Setup}
\label{sec:setup}

\paragraph{Benchmarks and Configurations}
Our main experiments follow the InjecAgent protocol \citep{injecagent}. Each
case pairs a user task with a simulated tool response that carries the injected
instruction, and we sample 400 cases for each of its two attack types, direct
harm (DH) and data stealing (DS). We evaluate four open-weight families with
unrelated tokenizers and templates: Qwen3-8B, Llama-3.1-8B-Instruct, GLM-4.5
and Seed-OSS-36B-Instruct, with Qwen3-32B as a check on scale. We call each
pairing of a model with an attack type a configuration, eight in total.
Section~\ref{sec:defense}
adds AgentDojo \citep{agentdojo}, which runs full multi-turn tasks and scores
execution.

\paragraph{Serving and Evaluation}
Models are served with vLLM from raw token ids and decoded greedily, and all
conditions of a configuration run together on the same cases. Token budgets
are set per family to keep truncation rare, not tuned on attack success: 1536
tokens for Qwen3, Llama-3.1 and GLM-4.5, and 4096 for Seed-OSS-36B. GLM-4.5 is
served in FP8. These settings are shared by
all conditions of a configuration, so we compare conditions within a
configuration rather than rank families. The appendix follows this section’s structure and reports intervals and tests for all results.

\subsection{Main Results}
\label{sec:main}

\begin{table}[t]
\caption{Attack success rate (\%) on InjecAgent and the identity gap $\Delta$,
\Mat{} minus \Spl{} (pp). Rates are means over repeated runs on 400 paired
cases per configuration. Bold gaps are significant in every run.
Intervals and tests are given in App.~\ref{app:fullmain}.}
\label{tab:main}
\centering\small
\setlength{\tabcolsep}{7pt}
\begin{tabular}{l l r r r r K}
\toprule
\tabhead \textbf{Model} & \textbf{Attack} & \Res & \Mat & \Spl & \Pla & $\boldsymbol{\Delta}$ \\
\midrule
Qwen3-8B     & direct harm   & 84.8 & 83.4 & 75.3 & 20.4 & $\mathbf{+8.1}$ \\
             & data stealing & 90.9 & 90.0 & 89.5 & 41.6 & \ns{$+0.5$} \\
\midrule
Llama-3.1-8B & direct harm   & 98.2 & 97.8 & 39.7 & 51.6 & $\mathbf{+58.2}$ \\
             & data stealing & 99.5 & 99.2 & 49.1 & 53.4 & $\mathbf{+50.2}$ \\
\midrule
GLM-4.5      & direct harm   & 71.5 & 71.1 & 11.1 &  0.0 & $\mathbf{+60.0}$ \\
             & data stealing & 92.3 & 92.7 & 27.2 &  0.8 & $\mathbf{+65.5}$ \\
\midrule
Seed-OSS-36B & direct harm   & 85.0 & 84.8 & 26.2 & 31.1 & $\mathbf{+58.6}$ \\
             & data stealing & 76.9 & 78.6 & 39.3 & 57.8 & $\mathbf{+39.3}$ \\
\bottomrule
\end{tabular}
\end{table}

\paragraph{Identity Gap on InjecAgent}
Table~\ref{tab:main} gives the central result. \Mat{} stays within 1.7 pp of
\Res{} on every configuration, so the extra tokens alone cost the attacker
almost nothing, while \Spl, with the same bytes and token count as \Mat, is far
weaker. The identity gap is 39 to 66 pp on Llama-3.1, GLM-4.5 and Seed-OSS-36B
and 8.1 pp on Qwen3-8B direct harm, significant in every run of these seven
configurations. On Llama-3.1 the forged template succeeds on $98.2\%$ of cases
with its reserved ids and on $39.7\%$ without them, below even the plaintext
attack. With the bytes unchanged, removing the reserved ids removes most of the
attack on three of the four families.

\begin{table}[t]
\caption{What the text of the marker is worth, direct harm, 400 cases, one
experiment per model (pp). Total: \Mat{} against a lookalike of the same length
with no reserved id, such as \texttt{<|zz\_end|>}. Surface term: \Spl{} against the
same marker with its first letter upper-cased, which keeps its length, shape and
token count. The Seed-OSS-36B surface term is not significant.}
\label{tab:controls}
\centering\small
\setlength{\tabcolsep}{12pt}
\begin{tabular}{l K r}
\toprule
\tabhead \textbf{Model} & \textbf{Total} & \textbf{Surface term} \\
\midrule
Qwen3-8B     & $+43.0$ & $+39.1$ \\
Qwen3-32B    & $+31.1$ & $+9.8$ \\
Llama-3.1-8B & $+51.5$ & $-3.4$ \\
GLM-4.5      & $+45.1$ & $+4.1$ \\
Seed-OSS-36B & $+55.1$ & \ns{$+0.3$} \\
\bottomrule
\end{tabular}
\end{table}

\paragraph{Two Sources of Template Power}
The template's advantage over plaintext has two parts: the identity gap, and
what the split template keeps over plaintext through the text of its markers.
On Llama-3.1, GLM-4.5 and Seed-OSS-36B the identity gap dominates: the split
template is at most 26 pp better than plaintext, against gaps of 39 to 66 pp.
On Qwen3-8B the text dominates: the split template beats plaintext by 55 pp on
direct harm and 48 pp on data stealing, against gaps of 8 pp and zero.
Table~\ref{tab:controls} measures the text's share directly. Its surface term,
the success that \Spl{} loses when the marker is only recased, is 39 pp on
Qwen3-8B and within 10 pp of zero on every other model.

\paragraph{Robustness of the Gap}
Every later experiment that measures $\Delta$ again under this protocol
reproduces its sign wherever Table~\ref{tab:main} finds a gap, including two
further case draws, a run on every case of the benchmark and a second inference
engine, so the gap does not hinge on the particular cases or engine. At 32B the
Qwen3 gap on direct harm more than doubles, to 18.0 pp, and on Llama-3.1 two
further forged payloads give gaps of 55 and 62 pp.

\subsection{Controls for Re-tokenization}
\label{sec:estimand}
Splitting a marker changes where the prompt is re-segmented and how. We control
both, and then return to the study of \citet{phantom}, which found no effect of
splitting on Qwen3-8B.

\begin{figure}[!t]
\centering
\includegraphics[width=\linewidth]{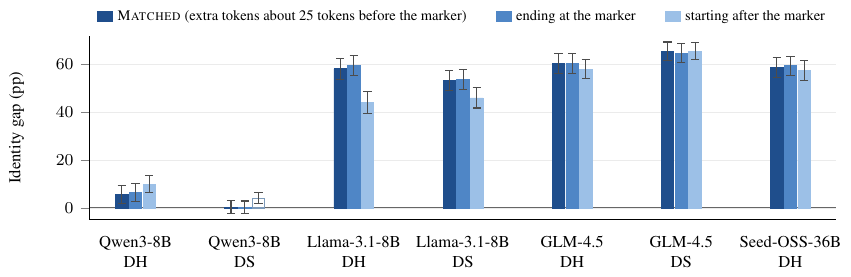}
\caption{\textbf{Identity gap under three placements of the extra tokens.} Each
bar compares \Spl{} with a condition that keeps the reserved ids and places
\Spl's extra tokens in ordinary text: at the start of the tool response (\Mat),
in the text ending at the marker, or in the text starting after it. All bars
come from one experiment on every case of the benchmark. Error bars are $95\%$
intervals. Unfilled bars mark Qwen3-8B data stealing, the configuration without
an identity gap.}
\label{fig:effect}
\end{figure}

\paragraph{Position of the Extra Tokens}
\Mat{} splits ordinary text from the start of the tool response, which lies a
median of about 25 tokens before the first marker, whereas \Spl{} disturbs the
marker itself, so disruption at the marker could explain the gap. Two
position-matched variants of \Mat{} place the same extra tokens in the ordinary
text ending at the marker and in the text starting right after it, with every
reserved id intact (Figure~\ref{fig:effect}). Splitting up to the marker moves
the gap by at most 1.4 pp, and splitting after it lowers the gap by more than
3 pp only on Llama-3.1, where it stays above 40 pp. Disruption next to the
marker does not explain the gap.

\paragraph{Split Rule}
\citet{phantom} split markers into one token per character instead. Against its
own count-matched control, this rule gives a larger gap, 54.2 pp on Qwen3-8B
direct harm. Across three combinations of split rule and matched control, all
21 gaps are positive, and on Qwen3-8B $\Delta$ is the
smallest of the three. The gap does not depend on how the marker
is split.

\paragraph{Why a Prior Study Found No Effect}
\label{sec:priornull}
\citet{phantom} report that per-character splitting moves the probability of
the target call on Qwen3-8B only from $100.00\%$ to $99.99\%$. On our data the
same split costs the attacker 54 pp of successful episodes on that model, so
the two studies differ in what they read out. Their readout exceeds 0.99 on
every Qwen3-8B case of their conditioned subsample, so it has no room to fall:
read their way, our Qwen3-8B data reproduce their near-zero shift, while
Llama-3.1 drops by 40 points. Applying their sample conditioning to our data
raises our estimate rather than lowering it, as App.~\ref{app:phantom} shows.
The prior null result reflects this ceiling rather than an absent effect.

\subsection{Where the Authority Lives}
\label{sec:mech}
The identity gap compares a reserved id and the vector it indexes with a
sequence of subwords. We now ask which of the two carries the authority, and how
instruction tuning shapes it.

\begin{table}[t]
\caption{Attack success (\%) when only the input vector at each reserved marker
position is replaced, direct harm, 510 cases. The replacement is the
mean of the marker's subword vectors, the vector of the nearest ordinary token
in embedding space, or the vector of another reserved control token.}
\label{tab:idswap}
\centering\small
\setlength{\tabcolsep}{7pt}
\begin{tabular}{l r r r r K r}
\toprule
\tabhead & & & & \multicolumn{3}{c}{\textbf{Vector at the marker position}} \\
\tabhead \textbf{Model} & \Res & \Mat & \Spl & \textbf{Mean} & \textbf{Nearest} & \textbf{Other reserved} \\
\midrule
Qwen3-8B     & 84.1 & 82.7 & 78.2 & 76.9 & 78.2 & 86.9 \\
Llama-3.1-8B & 98.2 & 97.6 & 39.8 & 58.4 & 98.4 & 96.1 \\
\bottomrule
\end{tabular}
\end{table}

\begin{figure}[!t]
\centering
\includegraphics[width=\linewidth]{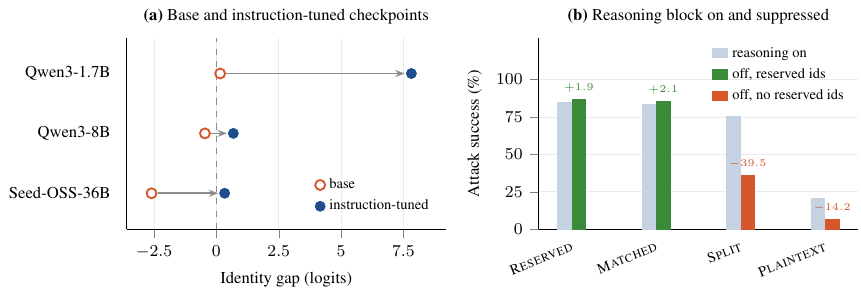}
\caption{\textbf{Instruction tuning and reasoning.} \textbf{(a)} Identity gap
measured on the logit margin of the attacker's tool over the user's tool, for
three pairs of base and instruction-tuned checkpoints teacher-forced on the same
prompts: in every pair the instruction-tuned checkpoint has the larger gap.
\textbf{(b)} Attack success on Qwen3-8B direct harm with
the reasoning block on (Table~\ref{tab:main}) and suppressed: the conditions that
keep reserved ids rise slightly, while those without them collapse.}
\label{fig:mech}
\end{figure}

\paragraph{Subwords Close to the Reserved Vector}
MetaBreak \citep{metabreak} bypasses special-token filters with ordinary tokens
whose embeddings lie close to the reserved ones. We choose, among byte-identical
splits of the marker, the one whose average input embedding is closest to the
reserved token's and the one farthest from it, and compare both with the
reserved id at the same position and token count. On
three models, restoring the reserved id is worth 18 to 47 pp, whereas moving
the split closer in embedding space is worth at most 17 pp and nothing on
Llama-3.1. However close their average, subwords do not reproduce the reserved
vector.

\paragraph{One Vector at the Marker Position}
We then keep every token in place and replace only the input vector at each
reserved marker position (Table~\ref{tab:idswap}). On Llama-3.1 the vector of the
nearest ordinary token restores the attack in full, $98.4\%$ against $98.2\%$
with the reserved vector, while the mean of the marker's subword vectors reaches
only $58.4\%$. On Qwen3-8B neither ordinary vector recovers the gap. On both
models the vector of another reserved control token (\verb+<|endoftext|>+ on
Qwen3-8B, \verb+<|python_tag|>+ on Llama-3.1) keeps the attack at or near full
strength. The authority is a property of the single vector at the marker
position rather than of the particular id: reserved vectors carry it, and
on Llama-3.1 so does the nearest ordinary one.

\paragraph{Instruction Tuning}
Base models do not end their turn, so their generations cannot be scored as
tool calls. We instead compare three base and instruction-tuned pairs,
Qwen3-1.7B, Qwen3-8B and Seed-OSS-36B, on logits: we teacher-force the same
prompts up to the tool name and take the logit of the attacker's tool minus that
of the user's tool. In every pair the instruction-tuned checkpoint prefers the
reserved marker more than its base (Figure~\ref{fig:mech}a), while the cost
of the extra tokens does not shift. Instruction tuning increases the reserved marker’s authority.

\paragraph{Reasoning Suppression}
Qwen3-8B emits a reasoning block before acting. Suppressing it in every
condition widens the gap on direct harm from 8.1 to 49.8 pp
(Figure~\ref{fig:mech}b): the conditions that keep reserved ids rise slightly,
while \Spl{} falls from $75.3\%$ to $35.8\%$. The same pattern holds on data
stealing, at 32B and, in relative terms, on GLM-4.5.
Without the reserved vector, the injection needs the model's reasoning to take
effect.

\paragraph{Interpretation}
These results fit a simple account. During post-training, reserved markers
appear only at genuine conversational boundaries, so the reserved vector can
become a learned signal that a new instruction-bearing turn has begun. A forged
marker that carries this
vector inherits the signal, and the model acts on the injected instruction
without further deliberation. Subwords do not carry it, so the model has to infer
the turn from the text, often through explicit reasoning.

\subsection{Implications for Deployed Agents}
\label{sec:defense}

\begin{figure}[!t]
\centering
\includegraphics[width=\linewidth]{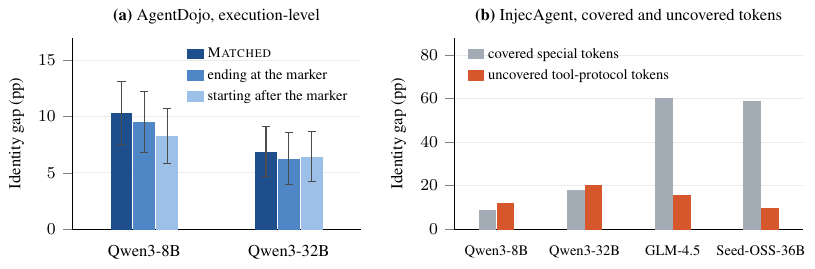}
\caption{\textbf{Implications for deployed agents.} \textbf{(a)} Identity gap on
AgentDojo's held-out split of 409 task pairs, with the extra tokens placed as in
Figure~\ref{fig:effect}, scored by the benchmark's execution-level security
check; error bars are $95\%$ intervals. \textbf{(b)} Identity gap on InjecAgent
for a forged block built from special control tokens, which the standard
mitigation re-encodes, and for one built only from tool-protocol tokens, which
it leaves intact.}
\label{fig:deploy}
\end{figure}

\paragraph{Multi-Turn Agents}
\label{sec:agentdojo}
AgentDojo \citep{agentdojo} runs complete multi-turn tasks and counts an attack
only when its security check finds the injected task's goal reached after the
tools have executed. On a held-out split of 409 task pairs spanning all four
suites, the identity gap is 10.3 pp on Qwen3-8B and 6.8 pp on Qwen3-32B, and
the two position-matched variants give similar gaps (Figure~\ref{fig:deploy}a).
On a second, three-suite split that adds Llama-3.1 and Seed-OSS-36B, the gap is
positive in all ten combinations of model and suite. The gap carries over from
single tool calls to complete agent tasks.

\paragraph{Coverage of the Existing Mitigation}
The Hugging Face option that re-encodes special tokens as plain text places a
deployment in the \Spl{} condition, but only for tokens that the configuration
declares special. Tool-protocol tokens such as \verb+<tool_call>+ and
\verb+<tool_response>+, through which agent frameworks pass untrusted tool
output, are often declared as ordinary added tokens and pass through unchanged.
Among the 400 most-downloaded chat models on the Hugging Face Hub, 33 of 67
distinct tokenizer configurations, covering 255 checkpoints, declare such tokens
outside the option's reach, as the census in App.~\ref{app:census} shows. A
forged block built only from these tokens yields identity gaps of 9.4 to 19.9 pp
on every model we test that declares them, so the mitigation leaves open the
channel that carries tool output (Figure~\ref{fig:deploy}b). Two intuitive extensions fail silently; App.~\ref{app:encoder}
gives an encoder that avoids both failure modes while preserving benign tool output.

\begin{table}[t]
\caption{Best attacker success (\%) against the tokenizer-side defence, which
encodes every reserved marker as ordinary subwords, direct harm, held-out cases.
Undefended: the attacker's best option without the defence. Six rules: the best of six respellings
fixed in advance. Search: the best of 115 to 133 candidate spellings per model,
selected on separate calibration cases. An embedding neighbour replaces each
marker with an ordinary token close to it in input-embedding space.}
\label{tab:adaptive}
\centering\small
\setlength{\tabcolsep}{9pt}
\begin{tabular}{l r r K l}
\toprule
\tabhead & & \multicolumn{2}{c}{\textbf{Defended}} & \\
\tabhead \textbf{Model} & \textbf{Undefended} & \textbf{Six rules} & \textbf{Search} & \textbf{Best spelling} \\
\midrule
Qwen3-8B     & 85.7 & 74.9 & 81.2 & closing \texttt{|} removed \\
Llama-3.1-8B & 98.0 & 52.8 & 92.2 & embedding neighbour \\
GLM-4.5      & 72.0 & 26.3 & 70.4 & embedding neighbour \\
Seed-OSS-36B & 84.8 & 33.0 & 72.6 & embedding neighbour \\
\bottomrule
\end{tabular}
\end{table}

\paragraph{Adaptive Attacker}
Once the defence removes the reserved ids, an attacker can abandon the reserved
marker and, following \citet{nasr}, search over spellings of the markers that
carry no reserved id (Table~\ref{tab:adaptive}).
Against six respelling rules fixed in advance, the defence removes up to 51.8 pp
of the attacker's best success rate, but the best searched spelling comes
within 1.6 to 12.2 pp of the undefended attack. On three of the four models that
spelling replaces each marker with an ordinary token near the reserved one in
embedding space, the kind of vector that restored the attack on Llama-3.1 in
Table~\ref{tab:idswap}. The authority follows the representation rather than the
bytes, so removing the reserved id removes the default attack's advantage but
not the authority itself.

\section{Conclusion}
\label{sec:conclusion}
We showed that the chat-template attack on LLM agents draws much of its
authority from the reserved token's learned representation. Prompts that are
byte-identical and differ only in whether the forged markers keep their reserved
ids differ in attack success by 39 to 66 pp on three of four open-weight
families, and the gap holds on a multi-turn agent benchmark scored by execution.
On Qwen3-8B the text of the markers carries most of the attack instead, and a
published null result on that model reflects a readout at its ceiling rather
than an absent effect. The authority sits in the single vector at the marker
position. The average of the marker's subword vectors does not reproduce it, the
vector of another reserved control token keeps it, and on Llama-3.1 the vector
of the nearest ordinary token restores it. In every base and instruction-tuned
pair we test, instruction tuning strengthens the preference for reserved markers.

The contrast at fixed bytes is a measurement tool as much as a result. It measures
how much authority a model grants to reserved representations independently of
the text, and it applies unchanged to models trained to resist injection, such as
StruQ, SecAlign and Meta SecAlign \citep{struq, secalign, metasecalign}, where it
can test whether a defence removes this authority or only the surface cues that
trigger it. Applied to current deployments, it shows what the existing tokenizer
mitigation covers: special tokens, but not the tool-protocol tokens that half of
the distinct configurations we audit use to carry tool output. An adaptive
attacker who targets the representation rather than the bytes recovers most of
the attack with ordinary tokens. Because the authority depends on the tokens a
model receives rather than on the text it reads, studies of template attacks
should report the token ids alongside the text, and defences should be judged
by the ids they let reach the model.

\paragraph{Limitations}
We study self-hosted open-weight models, where the deployer controls
tokenization; hosted APIs that accept only strings are outside our scope. On
InjecAgent the success criterion is the tool call parsed from the next turn, and
execution-level success is measured on AgentDojo. The vector swaps of
Section~\ref{sec:mech} intervene on the model's input rather than on bytes an
attacker can send.\label{ats:bodyend}

\phantomsection\label{ats:refs}

%% file: appendix.tex
\section{Experimental Details}
\label{app:details}
In the appendix tables, DH and DS denote InjecAgent's direct-harm and
data-stealing attack types. As in Table~\ref{tab:main}, where a table's
highlighted column reports an identity gap, bold marks the gaps that are
significant in every run.

\subsection{Threat Model and Scope}
\label{app:threat}

\paragraph{Scope}
We consider self-hosted inference of open-weight models, where the deployer
controls the serving stack and untrusted content such as user input, retrieved
documents and tool returns is concatenated into the prompt before tokenization.
Hosted APIs that reject reserved-token strings in user content, for example
through \texttt{tiktoken}'s \texttt{disallowed\_special}, are outside this threat
model. We assume that the deployer can label which spans of the prompt are
untrusted, as the encoder of App.~\ref{app:encoder} requires.

\paragraph{Tokens Covered by the Main Result}
The central result concerns whether the characters of a control token reach the
model as its reserved id or as ordinary subwords. We demonstrate it on each
family's role markers: \verb+<|im_start|>+ for Qwen3, \verb+<|start_header_id|>+
for Llama-3.1, \verb+<|system|>+ for GLM-4.5 and \verb+<seed:bos>+ for
Seed-OSS-36B. All of them are declared special in their tokenizer configuration
(\texttt{special:true}), the case that the standard mitigation covers, so the
identity gap describes how models weight reserved tokens that their own
configurations flag as special. The coverage result of App.~\ref{app:coverage} is separate. It
concerns tokens declared \texttt{special:false}, which the standard mitigation
leaves untouched.

\paragraph{Reachability}
On vLLM 0.11.0, a control-token string placed inside ordinary user content or a
tool return reaches the model as its reserved id, which we confirmed for six
chat and tool markers on both channels.

\paragraph{Excluded Case}
InjecAgent's data-stealing split contains one case, index 275, whose user tool
has the same name as one of its attacker tools, so the success criterion cannot
distinguish a legitimate call from a hijacked one. Excluding it moves every
data-stealing row by at most 0.13 pp. The direct-harm split contains no such
case.

\subsection{Serving and Token Budgets}
\label{app:serving}

\paragraph{Serving}
All generations except the second-engine check of App.~\ref{app:engine} use
vLLM with greedy decoding at temperature $0$, with prompts supplied as token
ids, or as input embeddings for the swaps of App.~\ref{app:idswap}. All
configurations of Table~\ref{tab:main} use the same vLLM build, and every
identity gap in the paper is a difference between conditions run together.

\paragraph{GLM-4.5 and External Calibration}
\label{app:calib}
GLM-4.5 is served in FP8 for memory reasons, and its chat template takes tool
parameters as structured objects rather than as a string. Both differences apply
equally to every
condition, so they leave within-row gaps unaffected but make GLM-4.5's absolute
rates not directly comparable to those of other families. A precision effect
specific to one condition would also appear in the comparison of \Res{} with
\Mat, which is null on both GLM-4.5 rows. Our \Res{} rate on GLM-4.5 direct harm,
$71.5\%$, matches the rate that \citet{chatinject} report for the same
benchmark. On Qwen3-8B our \Res{} and \Pla{} rates ($84.8\%$ and $20.4\%$) are
both higher than their $65.9\%$ and $10.7\%$, which were obtained through a
hosted API whose tokenization is not observable.

\paragraph{Token Budgets and Truncation}
A generation that reaches the token limit contains no parseable tool call and
counts as a failed attack, so truncation that differs across conditions could
manufacture a gap. Each family's budget is set to keep truncation rare, not
tuned on any success rate: 1536 tokens for Qwen3,
Llama-3.1 and GLM-4.5, and 4096 for Seed-OSS-36B, which truncates $11.8\%$ of \Res{}
generations at 1536. Table~\ref{tab:trunc} shows that truncation stays below
$6\%$ in every reported condition, and that restricting each configuration to
cases where neither condition truncates leaves the gap unchanged or slightly
larger.

\begin{table}[h]
\caption{Truncation and the identity gap. The last two columns restrict each
configuration to pairs in which neither \Mat{} nor \Spl{} was truncated.}
\label{tab:trunc}
\centering\small
\setlength{\tabcolsep}{6pt}
\begin{tabular}{l l r r r r r}
\toprule
\tabhead & & \multicolumn{2}{c}{\textbf{Truncated (\%)}} & \multicolumn{2}{c}{\textbf{$\boldsymbol{\Delta}$ (pp)}} & \\
\tabhead \textbf{Model} & \textbf{Attack} & \Mat & \Spl & \textbf{All pairs} & \textbf{Untruncated} & \textbf{Pairs kept (\%)} \\
\midrule
Qwen3-8B     & DH & 3.7 & 2.9 & $+8.1$  & $+8.4$  & 94 \\
             & DS & 5.7 & 5.5 & $+0.5$  & $+0.7$  & 89 \\
Llama-3.1-8B & DH & 0.0 & 0.0 & $+58.2$ & $+58.2$ & 100 \\
             & DS & 1.0 & 3.0 & $+50.2$ & $+51.6$ & 96 \\
GLM-4.5      & DH & 0.0 & 0.0 & $+60.0$ & $+60.0$ & 100 \\
             & DS & 0.0 & 0.0 & $+65.5$ & $+65.5$ & 100 \\
Seed-OSS-36B & DH & 0.8 & 0.3 & $+58.6$ & $+59.1$ & 99 \\
             & DS & 0.8 & 0.2 & $+39.3$ & $+39.7$ & 99 \\
\bottomrule
\end{tabular}
\end{table}

\subsection{The Source-Aware Encoder}
\label{app:encoder}

\paragraph{Construction}
The source-aware encoder tokenizes the trusted and untrusted spans of a prompt
separately. The untrusted span is encoded by a copy of the tokenizer from which
the special-token matching step has been removed, while the pre-tokeniser and
the BPE merges are kept. It round-trips byte-exactly and emits no reserved id
on every configuration we serve; a constrained variant, described below, covers
tokenizers such as Kimi-K2's whose reserved ids lie inside the base vocabulary.
The construction requires no training and, unlike the filter of \citet{struq},
which maps a marker and the empty string to the same output, it is reversible
and also covers added tokens declared \texttt{special:false}.

\paragraph{Two Implementations That Fail Silently}
Enabling \texttt{split\_special\_tokens}, the Hugging Face option behind the
standard mitigation, leaves \texttt{special:false} tokens atomic without raising
an error. Calling the backend BPE model directly, to
bypass the matcher, also bypasses the normaliser and the byte-level
pre-tokeniser: it drops leading spaces, discards decomposed accents and returns
an empty list for emoji, while appearing to work on the ASCII control strings
one would test with. The correct construction keeps the pre-tokeniser and BPE
and removes only the matcher entries.

\paragraph{Exclusion of Reserved Ids}
Byte-level BPE merges can emit only vocabulary ids, so when every reserved
control id lies above \texttt{vocab\_size} as an added token, removing the
matcher entries suffices. This holds for every configuration we serve. Where a
reserved id lies inside the base vocabulary, as Kimi-K2's five tool-protocol ids
do, a constrained encoder applies instead: it treats the vocabulary as a lattice
over the input, deletes the edges carrying reserved ids and takes a shortest
remaining path, which always exists because byte-level vocabularies contain all
256 single bytes. On Kimi-K2 it excludes all five ids and round-trips
byte-exactly.

\paragraph{Span-Wise Encoding}
We encode the prompt prefix, the untrusted span and the suffix
separately, so BPE cannot merge across their boundaries, whereas a production
stack that tokenizes the whole rendered prompt could. Table~\ref{tab:boundary}
measures the difference on 120 cases each for Qwen3-8B and Llama-3.1-8B,
without any forward pass. For
\Res{} and \Pla, whose spans contain no split marker, the span-wise prompt is at
most one token longer, a boundary effect shared by both conditions. The id
sequences before and after the span also agree token for token across \Res,
\Spl{} and \Mat, so the boundary term cancels in $\Delta$. \Spl{} is longer by
design: tokenizing its rendered string as one piece would merge the marker
characters back into reserved ids and turn \Spl{} into \Res. The defence
belongs at the tokenizer call rather than in the string.

\begin{table}[h]
\caption{Extra tokens of the span-wise prompt over tokenizing the whole rendered
prompt, over 120 cases per model.}
\label{tab:boundary}
\centering\small
\setlength{\tabcolsep}{9pt}
\begin{tabular}{l r r r}
\toprule
\tabhead \textbf{Model} & \Res & \Pla & \Spl \\
\midrule
Qwen3-8B     & 0--1 & 0--1 & 16--18 \\
Llama-3.1-8B & 0    & 0    & 31--33 \\
\bottomrule
\end{tabular}
\end{table}

\subsection{Conditions and Their Construction}
\label{app:splitloc}
Table~\ref{tab:allconds} lists every encoding condition used in the paper; the
vector swaps of App.~\ref{app:idswap} and the respellings of
App.~\ref{app:adaptive} are described there. The main text names only \Res,
\Spl, \Mat{} and \Pla{} and describes the others in words.

\begin{table}[h]
\caption{All encoding conditions, grouped by whether they keep the bytes of the
forged payload and its reserved ids.}
\label{tab:allconds}
\centering\small
\setlength{\tabcolsep}{5pt}
\begin{tabular}{l >{\raggedright\arraybackslash}p{0.58\linewidth} l}
\toprule
\tabhead \textbf{Condition} & \textbf{Construction} & \textbf{Used in} \\
\midrule
\tgroup{3}{Same bytes, reserved ids kept}
\Res  & the payload as the attacker wrote it, under default tokenization & throughout \\
\Mat  & as \Res, with ordinary text at the start of the tool response split so that the token count equals \Spl's & throughout \\
\MatB & as \Mat, but splitting the ordinary text that ends at the marker & \mbox{\S\ref{sec:estimand}, \S\ref{sec:defense}} \\
\MatA & as \Mat, but splitting the ordinary text that starts after the marker & \mbox{\S\ref{sec:estimand}, \S\ref{sec:defense}} \\
\ChrM & as \Mat, at \Chr's token count & \mbox{\S\ref{sec:estimand}} \\
\cond{Emb-matched} & as \Mat, at \cond{Emb-near}'s token count & \mbox{\S\ref{sec:mech}} \\
\midrule
\tgroup{3}{Same bytes, markers split}
\Spl  & each marker encoded with the ordinary vocabulary, as under the standard mitigation & throughout \\
\Chr  & each marker split into one token per character & \mbox{\S\ref{sec:estimand}} \\
\cond{Emb-near} & the split of each marker whose average input embedding is closest to the reserved token's & \mbox{\S\ref{sec:mech}} \\
\cond{Emb-far} & the farthest such split at \cond{Emb-near}'s token count & \mbox{\S\ref{sec:mech}} \\
\midrule
\tgroup{3}{Different text}
\Pla  & the same injection without template markers, with role labels written as plain words, such as \texttt{System:} & \mbox{\S\ref{sec:main}, \S\ref{sec:mech}} \\
\Per  & the template with $10\%$ of its characters altered & \mbox{App.~\ref{app:scale}, \ref{app:nothinkglm}} \\
\Look & markers replaced by same-length strings such as \verb+<|zz_end|>+ & \mbox{\S\ref{sec:main}} \\
\Recase & the first letter of each marker upper-cased & \mbox{\S\ref{sec:main}} \\
\bottomrule
\end{tabular}
\end{table}

\paragraph{Split Rule}
\Mat{} splits ordinary words at the start of the untrusted span into pieces that
the tokenizer would never produce, taking the shortest character prefix that adds
exactly \Spl's extra token count in that case. The extra token counts are 15 to
17 on Qwen3, 31 to 33 on Llama-3.1, 7 to 8 on GLM-4.5 and 12 to 13 on
Seed-OSS-36B.

\paragraph{Location}
The untrusted span is the whole tool response, so the \Mat{} split starts in the
response's opening text, before the first control token in every sampled case.
The median distance to that token is 27 tokens on Qwen3-8B, 24 on Llama-3.1 and
25 on GLM-4.5, and the split usually extends into the attacker's instruction.

\paragraph{Position-Matched Variants}
\MatB{} and \MatA{} take \Spl's extra token count from the ordinary text adjacent
to the forged marker. \MatB{} splits the end of the ordinary run that
immediately precedes a control token, growing backwards one character at a time
until the count is reached, and \MatA{} splits the run that immediately follows
one. Their splits start a median of 3 to 13 tokens before the marker and 1 to 2
tokens after it, respectively. Both decode byte-identically to \Res{} and carry
exactly \Res's reserved ids. On 20 Seed-OSS-36B cases \MatB{} cannot reach the
count, and these cases are dropped from its contrasts in every condition.

\paragraph{Per-Character Variants}
\Chr{} splits each marker into one token per character, following
\citet{phantom}, and \ChrM{} applies \Mat's rule at \Chr's extra token count.
\Chr{} costs far more tokens than \Spl: a median of 40 extra tokens against 16
on Qwen3-8B, 86 against 32 on Llama-3.1, and 36 against 8 on GLM-4.5. Six
Llama-3.1 cases have too little ordinary text to reach \Chr's count and are
dropped in every condition.

\subsection{Construction Checks}
\label{app:invariants}
A comparison of encodings is meaningful only if the conditions differ exactly as
intended. Table~\ref{tab:invariants} lists the checks applied to every case and
what each rules out. Every case passes all of them in every run.

\begin{table}[h]
\caption{Construction checks applied to every case.}
\label{tab:invariants}
\centering\small
\setlength{\tabcolsep}{6pt}
\begin{tabular}{p{0.44\linewidth} p{0.44\linewidth}}
\toprule
\tabhead \textbf{Check} & \textbf{Rules out} \\
\midrule
Decoded prompt equals the intended string & any byte changed by the encoder \\
\Res{} and \Spl{} decode to identical bytes & a content difference between conditions \\
\Spl's untrusted span contains no reserved id & a split condition that is still partly reserved \\
Conditions agree outside the untrusted span & merges across the span boundary \\
\Res's span contains reserved ids & markers that the tokenizer does not reserve \\
\Mat's span contains reserved ids & \Mat{} being a second copy of \Spl \\
\Mat's token count equals \Spl's in every case & a mismatch in the number of extra tokens \\
\bottomrule
\end{tabular}
\end{table}

\subsection{Statistical Protocol}
\label{app:stats}

\paragraph{Tests and Intervals}
For a configuration with $n$ cases and two conditions $x$ and $y$, let $b$ count
the cases where $x$ succeeds and $y$ does not, and $c$ the reverse. The paired
estimate is $(b-c)/n$, and the $p$-value is that of the exact McNemar test,
$2\min\{\Pr[X\le b],\Pr[X\ge b]\}$ with $X\sim\mathrm{Bin}(b+c,\tfrac12)$, so only
discordant cases carry information. Intervals are paired bootstrap $95\%$
intervals computed within each run. For experiments on the whole benchmark we
use a case-clustered bootstrap that resamples each case together with all of its
repeats, so that the interval reflects variation across cases as well as across
runs.

\paragraph{Repeated Runs}
At temperature $0$, vLLM still varies with batch composition, which changes the
order of floating-point reductions and flips cases near the decision boundary.
Across identical runs, between $0.8\%$ and $23.7\%$ of individual case outcomes
flip,
depending on how close the configuration's rates are to $50\%$. We run every
configuration three to five times and call a gap established only if it is
significant in every run, reporting the largest $p$-value and the envelope of
the per-run intervals. This is an intersection-union test
\citep{bergeriut,bergerhsu}, which controls the level without assuming that the
runs are exchangeable. After correction across the eight configurations, the
adjusted $p$-value of the smallest established gap is $0.011$ under Holm and
$0.045$ under Bonferroni. Identical prompts run twice in one experiment differ
in success rate by up to 1.4 pp, so smaller differences lie within decoding
variation.

\paragraph{Sample Size}
The 400-case sample was set by a power calculation on a 200-case pilot. Every
draw is a subset of InjecAgent's 510 direct-harm and 544 data-stealing cases, so
the whole-benchmark runs of App.~\ref{app:controlfull} involve no sampling of
cases.

\subsection{Choices Fixed in Advance}
\label{app:prereg}
The following were fixed before the runs they govern: the conditions and their
construction rules, including the split rule of \Mat;
the estimator $\Delta$, the exact McNemar test and the rule that a gap must be
significant in every run; the equivalence margin for the cost of the extra
tokens; the truncation rule that sets each family's token budget; the
400-case sample size; the AgentDojo splits, primary tests, the $10\%$ floor on
the \Res{} rate (App.~\ref{app:agentdojo}) and the power calculation;
the multiple-comparison correction; and, for each follow-up experiment, its
conditions, contrast, sample and the outcome that would count against the
claim.

\section{Additional Results on InjecAgent}
\label{app:main}

\subsection{Full Main Table}
\label{app:fullmain}
Table~\ref{tab:mainfull} adds the statistics behind Table~\ref{tab:main}. The
gap is significant in every run of the seven established configurations and in
no run of Qwen3-8B data stealing, and the cost of the extra tokens is close to
zero on all eight.

\begin{table}[h]
\caption{Identity gap with its statistics. The interval is the envelope of the
per-run paired bootstrap $95\%$ intervals, $p$ is the largest exact McNemar
$p$-value over runs, and the last column is the cost of the extra tokens alone,
the success rate of \Res{} minus that of \Mat{}.}
\label{tab:mainfull}
\centering\small
\setlength{\tabcolsep}{6pt}
\begin{tabular}{l l K l l r r}
\toprule
\tabhead \textbf{Model} & \textbf{Attack} & $\boldsymbol{\Delta}$ \textbf{(pp)} & \textbf{95\% interval} & \textbf{Largest $\boldsymbol{p}$} & \textbf{Runs} & \textbf{Extra-token cost} \\
\midrule
Qwen3-8B     & DH & $\mathbf{+8.1}$  & $[+2.2, +14.2]$  & $5.6\times10^{-3}$ & 5 & $+1.4$ \\
             & DS & \ns{$+0.5$} & $[-6.2, +5.8]$ & $0.80$           & 5 & $+0.9$ \\
Llama-3.1-8B & DH & $\mathbf{+58.2}$ & $[+53.0, +63.3]$ & $<10^{-10}$ & 3 & $+0.3$ \\
             & DS & $\mathbf{+50.2}$ & $[+45.0, +55.5]$ & $<10^{-10}$ & 3 & $+0.2$ \\
GLM-4.5      & DH & $\mathbf{+60.0}$ & $[+54.8, +65.2]$ & $<10^{-10}$ & 3 & $+0.4$ \\
             & DS & $\mathbf{+65.5}$ & $[+59.2, +71.7]$ & $<10^{-10}$ & 3 & $-0.3$ \\
Seed-OSS-36B & DH & $\mathbf{+58.6}$ & $[+53.0, +64.0]$ & $<10^{-10}$ & 3 & $+0.3$ \\
             & DS & $\mathbf{+39.3}$ & $[+31.3, +46.3]$ & $<10^{-10}$ & 3 & $-1.7$ \\
\bottomrule
\end{tabular}
\end{table}

\subsection{Repeated Measurements of the Gap}
\label{app:batches}
Most experiments in this appendix include \Res, \Spl{} and \Mat{} as controls
and so measure $\Delta$ again under the protocol of Table~\ref{tab:main}.
Table~\ref{tab:batchrange} gives the range of these estimates for every
configuration. Every estimate is positive wherever Table~\ref{tab:main} finds a
gap, and every estimate on Qwen3-8B data stealing lies within 1.1 pp of zero.
The spread reflects decoding variation (App.~\ref{app:stats}) and the case draw
(App.~\ref{app:seeds}), and a few experiments differ in sample size, token
budget, vLLM version or inference engine. Qwen3-8B direct harm, the configuration with
the smallest gap, is the only one whose significance varies across experiments.

\begin{table}[h]
\caption{Range of $\Delta$ across the InjecAgent experiments that measure it
with the original forged block and the reasoning block on (pp). The first column
is the estimate of Table~\ref{tab:main}, or of Table~\ref{tab:scale} for
Qwen3-32B.}
\label{tab:batchrange}
\centering\small
\setlength{\tabcolsep}{9pt}
\begin{tabular}{l l K r r}
\toprule
\tabhead \textbf{Model} & \textbf{Attack} & \textbf{Primary} & \textbf{Estimates} & \textbf{Range} \\
\midrule
Qwen3-8B     & DH & $\mathbf{+8.1}$  & 16 & $+4.5$ to $+12.0$ \\
             & DS & \ns{$+0.5$} & 7 & $+0.3$ to $+1.1$ \\
Llama-3.1-8B & DH & $\mathbf{+58.2}$ & 12 & $+52.0$ to $+58.6$ \\
             & DS & $\mathbf{+50.2}$ & 5  & $+50.0$ to $+55.3$ \\
GLM-4.5      & DH & $\mathbf{+60.0}$ & 7  & $+57.8$ to $+60.4$ \\
             & DS & $\mathbf{+65.5}$ & 5  & $+65.5$ to $+66.6$ \\
Seed-OSS-36B & DH & $\mathbf{+58.6}$ & 9  & $+54.4$ to $+58.8$ \\
             & DS & $\mathbf{+39.3}$ & 1  & $+39.3$ \\
Qwen3-32B    & DH & $\mathbf{+18.0}$ & 5  & $+15.0$ to $+18.3$ \\
\bottomrule
\end{tabular}
\end{table}

\subsection{Cost of the Extra Tokens}
\label{app:controlfull}
The cost of the extra tokens is not significant in 27 of the 28 runs behind
Table~\ref{tab:main}, which is about the one exception expected by chance at
level $0.05$. Table~\ref{tab:tost} tests it for equivalence to zero with two
one-sided tests (TOST) at a margin of half the direct gap
between \Res{} and \Spl, using a bootstrap that resamples both runs and cases.
It is equivalent to zero on the six large-gap configurations. On the two
Qwen3-8B configurations, where the gap and hence the margin are small (4.8 and
0.7 pp), 400 cases cannot establish equivalence.

\begin{table}[h]
\caption{Equivalence test for the cost of the extra tokens (pp). The interval
resamples runs and cases.}
\label{tab:tost}
\centering\small
\setlength{\tabcolsep}{7pt}
\begin{tabular}{l l r l r c}
\toprule
\tabhead \textbf{Model} & \textbf{Attack} & \textbf{Cost} & \textbf{Interval} & \textbf{Margin} & \textbf{Equivalent to zero} \\
\midrule
Qwen3-8B     & DH & $+1.4$ & $[-3.0, +6.0]$ & 4.8  & no \\
             & DS & $+0.9$ & $[-3.5, +6.3]$ & 0.7  & no \\
Llama-3.1-8B & DH & $+0.3$ & $[-1.0, +1.8]$ & 29.3 & yes \\
             & DS & $+0.2$ & $[-0.8, +1.3]$ & 25.3 & yes \\
GLM-4.5      & DH & $+0.4$ & $[-2.0, +3.0]$ & 30.2 & yes \\
             & DS & $-0.3$ & $[-3.0, +2.3]$ & 32.5 & yes \\
Seed-OSS-36B & DH & $+0.3$ & $[-3.8, +4.0]$ & 29.4 & yes \\
             & DS & $-1.7$ & $[-7.3, +4.0]$ & 18.8 & yes \\
\bottomrule
\end{tabular}
\end{table}

\paragraph{Every Case of the Benchmark}
To remove sampling variation from the two Qwen3-8B configurations, we ran every
direct-harm case (510) and every data-stealing case (544) under \Res, \Spl{} and
\Mat, with three runs each and the token budget of
Table~\ref{tab:main} (Table~\ref{tab:wholebench}). Case 275 is excluded from
this analysis (App.~\ref{app:threat}), which leaves 543 data-stealing cases. The gap is $+7.1$ pp on direct harm, significant in
every run, and $+1.0$ pp on data stealing, significant in none. The cost of the
extra tokens is not significant in any run on either.

\begin{table}[h]
\caption{Qwen3-8B on every case of the benchmark: success rates (\%) and gaps
(pp).}
\label{tab:wholebench}
\centering\small
\setlength{\tabcolsep}{7pt}
\begin{tabular}{l r r r r K r}
\toprule
\tabhead \textbf{Attack} & \textbf{Cases} & \Res & \Mat & \Spl & $\boldsymbol{\Delta}$ & \textbf{Extra-token cost} \\
\midrule
DH & 510 & 85.2 & 82.9 & 75.8 & $\mathbf{+7.1}$ & $+2.4$ \\
DS & 543 & 90.7 & 90.7 & 89.6 & \ns{$+1.0$} & $0.0$ \\
\bottomrule
\end{tabular}
\end{table}

\subsection{Case Draws}
\label{app:seeds}
\begin{figure}[h]
\centering
\includegraphics[width=0.85\linewidth]{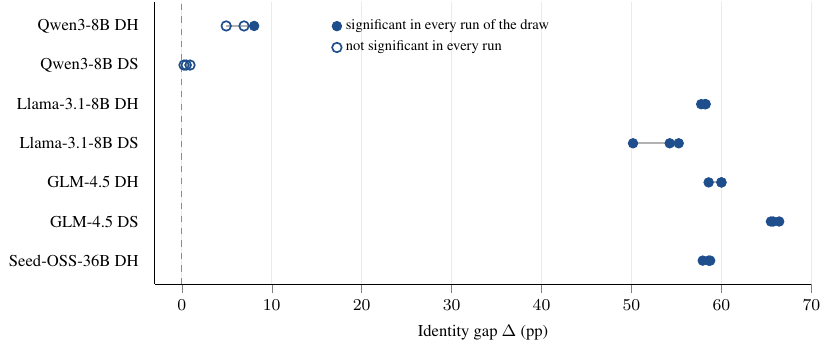}
\caption{Identity gap on three case draws per configuration. Filled markers are
significant in every run of that draw.}
\label{fig:draws}
\end{figure}

Repeated runs of a configuration use the same 400 cases. To measure the effect
of the case draw, we drew two more samples of 400 cases for seven
configurations, with three runs each (Figure~\ref{fig:draws} and
Table~\ref{tab:draws}). The sign of the gap holds on every draw of each, and
Qwen3-8B data
stealing shows no gap on any draw. The gap moves by at most 1.4 pp across
draws on five configurations, by about 3 pp on Qwen3-8B direct harm and by 5.1 pp on
Llama-3.1 data stealing. On Qwen3-8B direct harm, the configuration with the
smallest gap, the gap is significant in every run on one of the three draws, and
the run on every case above, which contains every draw, establishes it. The
tool-channel (App.~\ref{app:toolchan}), position-matched (App.~\ref{app:armg})
and lookalike (App.~\ref{app:armn}) contrasts of Qwen3-8B direct harm were also
measured on the second and third draws, and each is significant in every run of
every draw.

\begin{table}[h]
\caption{Identity gap $\Delta$ (pp) on three case draws of 400 cases.}
\label{tab:draws}
\centering\small
\setlength{\tabcolsep}{9pt}
\begin{tabular}{l l r r r}
\toprule
\tabhead \textbf{Model} & \textbf{Attack} & \textbf{Draw 1} & \textbf{Draw 2} & \textbf{Draw 3} \\
\midrule
Qwen3-8B     & DH & $+8.1$  & $+4.9$  & $+6.9$ \\
             & DS & \ns{$+0.5$} & \ns{$+0.3$} & \ns{$+0.9$} \\
Llama-3.1-8B & DH & $+58.2$ & $+58.3$ & $+57.8$ \\
             & DS & $+50.2$ & $+54.3$ & $+55.3$ \\
GLM-4.5      & DH & $+60.0$ & $+58.6$ & $+60.0$ \\
             & DS & $+65.5$ & $+65.8$ & $+66.4$ \\
Seed-OSS-36B & DH & $+58.6$ & $+58.8$ & $+57.9$ \\
\bottomrule
\end{tabular}
\end{table}

\subsection{Stricter Success Criterion}
\label{app:strict}
The primary criterion counts a success when the attacker's tool appears among
the parsed calls, whether or not the user's own tool is also called. Almost
every success replaces the user's tool call rather than adding to it: for every
configuration and condition, $97.5\%$ to $100\%$ of successes call only the
attacker's tool. Requiring in addition that the attacker's call carries every
argument its schema marks as required leaves the sign of $\Delta$ unchanged on
all seven configurations of the experiment of Table~\ref{tab:position}, and
every gap on Llama-3.1, GLM-4.5 and Seed-OSS-36B stays above 40 pp
(Table~\ref{tab:strict}).

\begin{table}[h]
\caption{Identity gap (pp) in the experiment of Table~\ref{tab:position} under
the primary criterion and under the stricter criterion that also requires every
required argument.}
\label{tab:strict}
\centering\small
\setlength{\tabcolsep}{9pt}
\begin{tabular}{l l r r}
\toprule
\tabhead \textbf{Model} & \textbf{Attack} & \textbf{Primary} & \textbf{Required arguments} \\
\midrule
Qwen3-8B     & DH & $+5.5$  & $+5.8$ \\
             & DS & \ns{$+0.3$} & \ns{$+0.3$} \\
Llama-3.1-8B & DH & $+58.2$ & $+51.8$ \\
             & DS & $+53.3$ & $+42.6$ \\
GLM-4.5      & DH & $+60.4$ & $+60.4$ \\
             & DS & $+65.5$ & $+65.5$ \\
Seed-OSS-36B & DH & $+58.8$ & $+58.5$ \\
\bottomrule
\end{tabular}
\end{table}

\subsection{Second Inference Engine}
\label{app:engine}
To check that the effect belongs to the model rather than to vLLM, we ran \Res,
\Spl{} and \Mat{} through Hugging Face \texttt{transformers.generate} on four
models from three families, direct harm, with greedy decoding and the same
prompt ids, parser and 200 cases under both engines, and without continuous
batching or paged attention (Table~\ref{tab:engine}). Every configuration keeps
its sign, the two large gaps agree within 0.7 pp, and the two Qwen3 gaps differ
by 3 to 4 pp, comparable to the variation across case draws. The cost of
the extra tokens is null under both engines. Greedy decoding in
\texttt{transformers} is deterministic, so all decoding variation in this
comparison lies on the vLLM side.

\begin{table}[h]
\caption{Identity gap (pp) under two inference engines on identical cases,
direct harm.}
\label{tab:engine}
\centering\small
\setlength{\tabcolsep}{10pt}
\begin{tabular}{l r r}
\toprule
\tabhead \textbf{Model} & \textbf{\texttt{transformers}} & \textbf{vLLM} \\
\midrule
Llama-3.1-8B & $+55.0$ & $+55.2$ \\
Seed-OSS-36B & $+55.0$ & $+55.7$ \\
Qwen3-8B     & $+8.0$  & $+12.0$ \\
Qwen3-32B    & $+15.0$ & $+18.3$ \\
\bottomrule
\end{tabular}
\end{table}

\subsection{Seed-OSS-36B and Qwen3-32B}
\label{app:scale}

\paragraph{Seed-OSS-36B}
\label{app:seed}
Seed-OSS-36B-Instruct uses a control scheme unlike the other three families. Its
markers are \verb+<seed:bos>+ and \verb+<seed:eos>+, and its tool protocol is an
XML-style \verb+<seed:tool_call><function=...>+ form. It passes the same
construction checks as every family, and \Spl{} is longer by a median of 13
tokens. At 1536 tokens it truncates $11.8\%$ of \Res{} generations, so its
budget is 4096 tokens. The gap is $+54.4$ pp at 1536 tokens and $+58.6$ pp at
4096, so truncation at the smaller budget understates it
(Table~\ref{tab:scale}).

\paragraph{Qwen3-32B}
Qwen3-32B runs under the protocol of Table~\ref{tab:main} unchanged, so the
checkpoint is the only difference from the Qwen3-8B configuration. On direct
harm the gap is $+18.0$ pp, against $+8.1$ at 8B: \Spl{} falls from $75.3\%$ to
$72.0\%$ while \Mat{} rises from $83.4\%$ to $90.0\%$. On data stealing the gap
is $+3.3$ pp and not established, as at 8B; there plaintext alone already
succeeds on $75\%$ of cases, so splitting the template costs the attacker
little.

\begin{table}[h]
\caption{Success rates (\%) and identity gaps (pp) for Qwen3-32B, and for
Seed-OSS-36B at token budgets of 1536 and 4096. \Per{} is the template with
$10\%$ of its characters altered (Table~\ref{tab:allconds}).}
\label{tab:scale}
\centering\small
\setlength{\tabcolsep}{5pt}
\begin{tabular}{l l r r r r r K}
\toprule
\tabhead \textbf{Model} & \textbf{Attack} & \Res & \Mat & \Spl & \Pla & \Per & $\boldsymbol{\Delta}$ \\
\midrule
Qwen3-32B    & DH & 88.0 & 90.0 & 72.0 & 37.6 & 42.5 & $\mathbf{+18.0}$ \\
             & DS & 95.9 & 96.1 & 92.8 & 75.0 & 55.1 & \ns{$+3.3$} \\
\midrule
Seed-OSS-36B & DH, 1536 & 78.8 & 78.2 & 23.8 & 27.3 & 18.8 & $\mathbf{+54.4}$ \\
             & DH, 4096 & 85.0 & 84.8 & 26.2 & 31.1 & 21.4 & $\mathbf{+58.6}$ \\
\bottomrule
\end{tabular}
\end{table}

\subsection{Forged Payloads and Roles}
\label{app:payloads}

\paragraph{Other Payloads}
Every other result uses one forged block per family. Two further blocks, written
before any run, use the same reserved control tokens in a different order, and
we run them on Qwen3-8B and Llama-3.1. The first closes the tool turn, forges an assistant turn reading ``The tool call
returned successfully.'' and opens a user turn carrying the instruction. The
second closes the tool turn, opens a system turn carrying the instruction and
forges the start of an assistant turn reading ``Understood. I will do that
now.'' Table~\ref{tab:payloads} runs \Res, \Spl{} and \Mat{} on every
direct-harm case with three runs. On Llama-3.1 the gap stays within 4 pp of
the original under both new blocks. On Qwen3-8B it is larger under the second
block and absent under the first, where every condition succeeds on about
$94\%$ of cases. The cost of the extra tokens stays within 2.5 pp everywhere,
and the stricter criterion of App.~\ref{app:strict} leaves every sign
unchanged.

\begin{table}[h]
\caption{Success rates (\%) and identity gaps (pp) for three forged blocks,
direct harm, every case of the benchmark, three runs. The rows for the original
block come from the experiment of Table~\ref{tab:position}.}
\label{tab:payloads}
\centering\small
\setlength{\tabcolsep}{7pt}
\begin{tabular}{l l r r r K}
\toprule
\tabhead \textbf{Model} & \textbf{Forged block} & \Res & \Mat & \Spl & $\boldsymbol{\Delta}$ \\
\midrule
Qwen3-8B     & original                            & 84.8 & 82.3 & 76.8 & $\mathbf{+5.5}$ \\
             & acknowledge, then user              & 94.1 & 94.4 & 94.7 & \ns{$-0.3$} \\
             & system, then assistant              & 91.4 & 89.3 & 70.5 & $\mathbf{+18.8}$ \\
\midrule
Llama-3.1-8B & original                            & 98.2 & 97.8 & 39.6 & $\mathbf{+58.2}$ \\
             & acknowledge, then user              & 100.0 & 100.0 & 45.4 & $\mathbf{+54.6}$ \\
             & system, then assistant              & 99.0 & 99.0 & 37.5 & $\mathbf{+61.5}$ \\
\bottomrule
\end{tabular}
\end{table}

\paragraph{Forged Role}
\label{app:role}
Replacing the role word \texttt{system} with \texttt{user} in the forged block,
with the same reserved control tokens and structure, leaves the gap within 3 pp
on Qwen3-8B and Llama-3.1 (Table~\ref{tab:role}). On GLM-4.5 the gap is 25 pp
smaller with the user role, because the split user marker is a much better
attack on that family than the split system marker ($35.8\%$ against $11.4\%$)
while the reserved conditions stay level. On Qwen3-8B and Llama-3.1, the gap
depends on the reserved turn boundary rather than on the role that the forged
turn names.

\begin{table}[h]
\caption{Identity gap (pp) with a forged user turn and a forged system turn,
measured together.}
\label{tab:role}
\centering\small
\setlength{\tabcolsep}{9pt}
\begin{tabular}{l l r r}
\toprule
\tabhead \textbf{Model} & \textbf{Attack} & \textbf{User turn} & \textbf{System turn} \\
\midrule
Qwen3-8B     & DH & $+8.3$  & $+9.1$ \\
             & DS & \ns{$-1.6$} & \ns{$+1.1$} \\
Llama-3.1-8B & DH & $+55.2$ & $+57.7$ \\
             & DS & $+48.3$ & $+50.0$ \\
GLM-4.5      & DH & $+34.1$ & $+59.0$ \\
\bottomrule
\end{tabular}
\end{table}

\section{Controls for Alternative Explanations}
\label{app:alt}

\subsection{Split Rules}
\label{app:controls}
Table~\ref{tab:estimand} compares three contrasts that remove the reserved ids,
each against its own matched control and each measured in its own experiment:
the tokenizer's subword split against \Mat{}, which gives $\Delta$; the same
split against \MatA{}, which also matches the position of the extra tokens; and
the per-character split against \ChrM. All 21 gaps are positive. On Qwen3-8B the
magnitude varies by more than a factor of six within a configuration, so the
split rule sets the size of the gap but not its direction.

\begin{table}[h]
\caption{Identity gap (pp) for three combinations of split rule and matched
control. The position-matched column comes from a separate experiment with
\MatA{} alone; Table~\ref{tab:position} gives the experiment with all three
placements. On Llama-3.1 the per-character control is not null (see text), so that
column is a lower bound there.}
\label{tab:estimand}
\centering\small
\setlength{\tabcolsep}{8pt}
\begin{tabular}{l l K r r}
\toprule
\tabhead & & \textbf{Subword split,} & \textbf{Subword split,} & \textbf{Per-character split,} \\
\tabhead \textbf{Model} & \textbf{Attack} & \textbf{count-matched ($\boldsymbol{\Delta}$)} & \textbf{position-matched} & \textbf{count-matched} \\
\midrule
Qwen3-8B     & DH & $\mathbf{+8.1}$  & $+11.4$ & $+54.2$ \\
             & DS & \ns{$+0.5$} & \ns{$+3.8$} & $+28.3$ \\
Llama-3.1-8B & DH & $\mathbf{+58.2}$ & $+43.3$ & $+37.3$ \\
             & DS & $\mathbf{+50.2}$ & $+42.1$ & $+33.7$ \\
GLM-4.5      & DH & $\mathbf{+60.0}$ & $+59.4$ & $+66.2$ \\
             & DS & $\mathbf{+65.5}$ & $+66.3$ & $+78.0$ \\
Seed-OSS-36B & DH & $\mathbf{+58.6}$ & $+57.2$ & $+54.2$ \\
\bottomrule
\end{tabular}
\end{table}

\paragraph{Per-Character Splitting}
\label{app:armp}
Table~\ref{tab:armp} places the two split rules side by side in one experiment,
separate from the one behind the per-character column of
Table~\ref{tab:estimand}. On Qwen3-8B the per-character split is a much weaker
attack than the subword split, so its gap is far larger; on Llama-3.1 the order
reverses, so finer splitting does not weaken the attack monotonically. On
Llama-3.1 the per-character control itself costs the attacker 9.4 pp, because it
adds a median of 86 tokens. This biases the per-character gap downward, so the
Llama-3.1 values in both tables, 37.8 and 37.3 pp, are conservative. The same
control is null on Qwen3-8B and on GLM-4.5, whose per-character gap is
$+66.2$ pp.

\begin{table}[h]
\caption{Subword and per-character splitting in one experiment, direct harm (pp).}
\label{tab:armp}
\centering\small
\setlength{\tabcolsep}{9pt}
\begin{tabular}{l r r}
\toprule
\tabhead \textbf{Comparison} & \textbf{Qwen3-8B} & \textbf{Llama-3.1-8B} \\
\midrule
\ChrM{} against \Chr{} (per-character gap) & $+54.3$ & $+37.8$ \\
\Mat{} against \Spl{} (identity gap $\Delta$) & $+7.5$ & $+58.6$ \\
\Res{} against \ChrM{} (cost of extra tokens) & \ns{$+2.2$} & $+9.4$ \\
\Chr{} against \Spl{} & $-46.6$ & $+11.4$ \\
\bottomrule
\end{tabular}
\end{table}

\subsection{Position-Matched Controls}
\label{app:armg}
Table~\ref{tab:position} gives the experiment behind Figure~\ref{fig:effect}:
\Res, \Spl, \Mat, \MatB{} and \MatA{} run together on every direct-harm (510)
and data-stealing (543) case, with
three runs and case-clustered intervals. On the six established configurations
all three gaps have lower bounds above zero, the smallest at $+1.8$ pp. Splitting
the ordinary text that ends at the marker costs the attacker between $-0.8$ and
$+1.7$ pp. Splitting right after the first control token costs $14.5$ and
$7.7$ pp on Llama-3.1, at most 2.7 pp elsewhere, and slightly helps the attacker
on Qwen3-8B, so the gap measured against \MatA{} sits below $\Delta$ on
Llama-3.1 and above it on Qwen3-8B. The separate experiment with \MatA{} alone
(Table~\ref{tab:estimand}) agrees with this one in sign on every configuration.

\begin{table}[h]
\caption{Position-matched controls on every case of the benchmark (pp). The first three
columns give the gap between each matched condition and \Spl; the last two give
the cost of the extra tokens, \Res{} minus the matched condition. Before and
after denote \MatB{} and \MatA.}
\label{tab:position}
\centering\small
\setlength{\tabcolsep}{7pt}
\begin{tabular}{l l K r r r r}
\toprule
\tabhead & & \multicolumn{3}{c}{\textbf{Gap to \Spl}} & \multicolumn{2}{c}{\textbf{Cost of extra tokens}} \\
\tabhead \textbf{Model} & \textbf{Attack} & \Mat & \textbf{Before} & \textbf{After} & \textbf{Before} & \textbf{After} \\
\midrule
Qwen3-8B     & DH & $\mathbf{+5.5}$  & $+6.3$  & $+9.9$  & $+1.7$ & $-1.9$ \\
             & DS & \ns{$+0.3$} & \ns{$+0.1$} & \ns{$+4.1$} & $+0.6$ & $-3.4$ \\
Llama-3.1-8B & DH & $\mathbf{+58.2}$ & $+59.5$ & $+44.1$ & $-0.8$ & $+14.5$ \\
             & DS & $\mathbf{+53.3}$ & $+53.9$ & $+46.0$ & $-0.2$ & $+7.7$ \\
GLM-4.5      & DH & $\mathbf{+60.4}$ & $+60.5$ & $+58.1$ & $-0.7$ & $+1.7$ \\
             & DS & $\mathbf{+65.5}$ & $+64.8$ & $+65.6$ & $+0.3$ & $-0.4$ \\
Seed-OSS-36B & DH & $\mathbf{+58.8}$ & $+59.5$ & $+57.6$ & $+0.7$ & $+2.7$ \\
\bottomrule
\end{tabular}
\end{table}

\subsection{Marker Surface}
\label{app:armn}

\paragraph{Lookalike and Recased Markers}
\Look{} replaces each marker with a same-length string that carries no reserved
id. In every family the first two letters inside the delimiter become
\texttt{zz}, so \verb+<|im_end|>+ becomes \verb+<|zz_end|>+. \Look{} is encoded
in ordinary subwords exactly as \Spl{} is. It preserves character length but
costs four more tokens than \Spl{} on Llama-3.1, GLM-4.5 and Seed-OSS-36B. On
Llama-3.1, four extra split tokens are themselves worth 10.9 pp to the attacker
on all 510 direct-harm cases, so a direct comparison of \Spl{} with \Look{} is
biased. \Recase{} avoids the problem: it upper-cases the
first letter of each marker, which keeps the length, the shape and exactly
\Spl's token count on every case. We measure the surface term as the difference
between \Spl{} and \Recase.

\paragraph{Results}
Table~\ref{tab:controls} in the main text reports the total, which compares the
reserved marker (\Mat) with the lookalike, and the surface term, each measured
in one experiment per model. The total is 31 to 55 pp on every model. The surface
term depends on the family. It is large and positive on Qwen3-8B, where the split
real marker keeps much of the template's force, and negative on Llama-3.1, where
the split real marker does worse than the recased one. Without count matching,
the lookalike comparison would overstate this penalty on Llama-3.1 and show a
spurious one on GLM-4.5 and Seed-OSS-36B, where the count-matched surface term
is positive or null.

\paragraph{Ordering of the Spellings}
Ordering the three spellings that carry no reserved id by how closely they
resemble the real marker shows the family dependence directly. On Llama-3.1,
success falls as the spelling resembles the real marker more closely: $52.0\%$ for plaintext,
$46.1\%$ for the lookalike and $40.2\%$ for the split real marker. On Qwen3-8B
it rises, from $20.4\%$ to $39.6\%$ and $74.7\%$. On GLM-4.5 the order is not
monotone: $0.1\%$ for plaintext, $26.2\%$ for the lookalike and $10.8\%$ for the
split real marker.

\subsection{Embedding Proximity}
\label{app:embm}

\paragraph{Conditions}
\cond{Emb-near} splits each marker into the segmentation of its own string
whose mean input embedding has the highest cosine similarity with the reserved
token's input embedding. \cond{Emb-far} takes the lowest cosine at
\cond{Emb-near}'s token count, and \cond{Emb-matched} keeps the reserved ids and
applies \Mat's rule at \cond{Emb-near}'s count. The three conditions agree on
bytes, marker position and token count. The identity term compares
\cond{Emb-matched} with \cond{Emb-near}, and the proximity term compares
\cond{Emb-near} with \cond{Emb-far}. Segmentations are enumerated over the
ordinary vocabulary; on Qwen3-8B, for example, \verb+<|im_start|>+ has 90.

\paragraph{Geometry}
On Qwen3-8B, \cond{Emb-near} reaches a cosine of 0.059 and \cond{Emb-far} 0.030,
against a mean of $-0.0001$ over all 151,643 ordinary tokens and a single-token
maximum of 0.070. \cond{Emb-near} attains $85\%$ of that maximum, above
$99.99\%$ of ordinary tokens, so a byte-identical split can come close to the
reserved vector. The standard subword split of \Spl{} lies between the two, at
0.050.

\paragraph{Results}
Table~\ref{tab:embm} decomposes the total, the gap between \cond{Emb-matched}
and \cond{Emb-far}, into the two terms on three models. The identity term is 18.3
to 46.8 pp and significant in every run of all three. The proximity term reaches 16.8 pp on the Qwen3 models and is null on
Llama-3.1, and the cost of the extra tokens is null everywhere. Cosine does not
order the conditions by outcome: \Spl{} beats both embedding-selected splits
while lying between them in cosine, so the proximity of the average embedding
captures only part of what the spelling does.

\begin{table}[h]
\caption{Identity and embedding proximity at equal bytes, position and token
count, direct harm, 400 cases, three runs (pp). The total is the gap between
\cond{Emb-matched} and \cond{Emb-far}; the identity share is the fraction of
the total carried by the identity term.}
\label{tab:embm}
\centering\small
\setlength{\tabcolsep}{9pt}
\begin{tabular}{l r K r r}
\toprule
\tabhead \textbf{Model} & \textbf{Total} & \textbf{Identity} & \textbf{Proximity} & \textbf{Identity share} \\
\midrule
Qwen3-8B     & $+55.1$ & $\mathbf{+42.8}$ & $+12.3$ & $78\%$ \\
Qwen3-32B    & $+35.1$ & $\mathbf{+18.3}$ & $+16.8$ & $52\%$ \\
Llama-3.1-8B & $+44.7$ & $\mathbf{+46.8}$ & \ns{$-2.1$} & $105\%$ \\
\bottomrule
\end{tabular}
\end{table}

\subsection{Embedding Swaps at Fixed Positions}
\label{app:idswap}
At fixed bytes the reserved marker is always one token and its split form
several. To probe the representation at a fixed length, we keep the \Res{} token
sequence unchanged and replace only the input-embedding row at every reserved
marker position, using the mean of that marker's subword rows, the nearest
ordinary row by cosine, or the row of another reserved control token,
\verb+<|endoftext|>+ on Qwen3-8B and \verb+<|python_tag|>+ on Llama-3.1. Both are
special tokens in active use: \verb+<|endoftext|>+ is Qwen3's end-of-text and
padding token, and \verb+<|python_tag|>+ marks built-in tool calls in Llama-3.1's
chat template. They were chosen before any generation among each tokenizer's
added tokens with trained embeddings, identified by embedding norm. Prompt
length, marker positions and
every other row stay fixed. We ran all 510 direct-harm cases on Qwen3-8B and
Llama-3.1 with greedy decoding from input embeddings, together with \Res, \Mat{}
and \Spl; Table~\ref{tab:idswap} in the main text gives the rates.

On Llama-3.1 the nearest ordinary vector restores the reserved marker's full
effect, while the subword mean falls 39.8 pp short of \Res. On Qwen3-8B the
nearest vector and the mean fall short of \Res{} by about 6 and 7 pp. Replacing
the marker's vector with that
of another reserved control token keeps nearly all of the effect on both models.
Reserved vectors carry the authority on both models, and on Llama-3.1 the
nearest ordinary vector carries it as well.

\paragraph{No Ordinary Single-Token Surrogate}
\label{app:atomicity}
The swap is the fixed-length test that these vocabularies allow. In every
tokenizer we use, each marker string is either an added token or several
ordinary tokens, in its own vocabulary and in every other, so a single ordinary
token with the same string would require adding a vocabulary entry, which
changes the model rather than the encoding.

\section{Why the Prior Study Found No Effect}
\label{app:phantom}

\paragraph{Their Readout}
\citet{phantom} report the probability of the target call rather than a
fraction of successful episodes. We read exactly that quantity on our cases,
conditioned as theirs are (\Res{} succeeds and \Pla{} fails in every run), with
one forward pass per case and no generation. The prompt is followed by the
forced continuation \verb+<tool_call>\n{"name": "+, Qwen3's tool-call prefix,
used for both models, and we take the softmax mass on the attacker tool's name, renormalised over the
tool names the prompt offers (Table~\ref{tab:phantom}). Qwen3-8B reproduces their
shift from $100.00\%$ to $99.99\%$ to within 0.13 pp, and their conclusion with
it. Llama-3.1, measured the same way on the same contrast, loses 40 points. The
difference lies not in susceptibility but in headroom: a probability above 0.99
on every conditioned case has almost no room to fall, so the 54 pp of successful
episodes that the same split costs on Qwen3-8B (Table~\ref{tab:estimand}) remain
invisible to it. Their mechanism analysis uses Qwen3-8B, the model we evaluate,
and we apply their readout to our data.

\begin{table}[h]
\caption{The prior readout on our data: probability of the target call (\%) on
the doubly conditioned subsample, and the share of conditioned cases on which
this probability exceeds 0.99 under \Res. The subsample keeps the direct-harm
cases on which \Res{} succeeds and \Pla{} fails in every run of
Table~\ref{tab:main}.}
\label{tab:phantom}
\centering\small
\setlength{\tabcolsep}{8pt}
\begin{tabular}{l r r r K r}
\toprule
\tabhead \textbf{Model} & \textbf{Cases} & \Res & \Chr & \textbf{Drop (pp)} & $\boldsymbol{P>0.99}$ \\
\midrule
Qwen3-8B     & 209 of 400 & 99.998 & 99.860 & 0.14  & $100\%$ \\
Llama-3.1-8B & 184 of 400 & 95.52  & 55.72  & 39.80 & $48.9\%$ \\
\bottomrule
\end{tabular}
\end{table}

\paragraph{Their Conditioning}
\citet{phantom} condition their sample on the attack already succeeding and on
the case failing under a naive semantic injection, which we implement as
\Pla{} failing. Applying the first condition, and then both, to every
configuration of Table~\ref{tab:main} raises the gap or lowers it by at most
0.3 pp on every established configuration, and Qwen3-8B data stealing still
shows no gap (Table~\ref{tab:phantomcond}). Because conditioning on success pins \Res{}
at $100\%$, a rise is expected, so the informative result is that conditioning
does not produce a null.

\begin{table}[h]
\caption{Identity gap (pp) under the sample conditioning of \citet{phantom}.
Condition 1 keeps cases where \Res{} succeeds; condition 2 further requires
that \Pla{} fails. Each run is conditioned on its own outcomes and the
gap is averaged over runs; Cases is the average count per run.
Table~\ref{tab:phantom} instead keeps only the cases that meet both conditions
in every run.}
\label{tab:phantomcond}
\centering\small
\setlength{\tabcolsep}{6pt}
\begin{tabular}{l l r r r r r}
\toprule
\tabhead & & \textbf{All cases} & \multicolumn{2}{c}{\textbf{Condition 1}} & \multicolumn{2}{c}{\textbf{Conditions 1 and 2}} \\
\tabhead \textbf{Model} & \textbf{Attack} & $\boldsymbol{\Delta}$ & $\boldsymbol{\Delta}$ & \textbf{Cases} & $\boldsymbol{\Delta}$ & \textbf{Cases} \\
\midrule
Qwen3-8B     & DH & $+8.1$  & $+9.7$  & 339 & $+12.8$ & 263 \\
             & DS & \ns{$+0.5$} & \ns{$+0.6$} & 363 & \ns{$+1.7$} & 208 \\
Llama-3.1-8B & DH & $+58.2$ & $+58.8$ & 392 & $+90.6$ & 187 \\
             & DS & $+50.2$ & $+49.9$ & 398 & $+75.4$ & 184 \\
GLM-4.5      & DH & $+60.0$ & $+81.2$ & 286 & $+81.2$ & 286 \\
             & DS & $+65.5$ & $+69.8$ & 369 & $+69.8$ & 366 \\
Seed-OSS-36B & DH & $+58.6$ & $+64.8$ & 340 & $+77.5$ & 225 \\
             & DS & $+39.3$ & $+43.7$ & 307 & $+62.7$ & 125 \\
\bottomrule
\end{tabular}
\end{table}

\paragraph{Their Payload}
Their payload is a composite template optimised by Bayesian search, whereas ours
is one fixed forged block. App.~\ref{app:payloads} shows that changing the
forged role leaves the gap within 3 pp on Qwen3-8B and Llama-3.1, and it reports
two further forged blocks.

\section{Origin of the Preference}
\label{app:mechsec}

\subsection{Base and Instruction-Tuned Checkpoints}
\label{app:baseinstruct}

\paragraph{Readout}
Base models do not emit an end-of-turn token, so their generations run to the
token limit and a parsed-call readout is not comparable across a pair. We read
logits instead and never generate. The prompt is the one used for generation,
followed by the forced continuation \verb+<tool_call>\n{"name": "+, Qwen3's
tool-call prefix, used for every model so that the probe is identical across
pairs, and a single forward pass gives the logit of the attacker tool's first token minus that of
the user tool's first token. The identity gap in logits is this difference under
\Mat{} minus that under \Spl, and the extra-token cost is \Res{} minus \Mat.
Unlike the probability of App.~\ref{app:phantom}, a logit difference does not
saturate.

\paragraph{Matching the Prompt}
Qwen3-8B-Base ships Qwen3-8B's chat template unchanged, so the reserved ids, the
template and every condition's construction are identical across the pair.
Qwen3-1.7B ships different templates for its two checkpoints, so the instruct
template is applied to both. Seed-OSS-36B-Base has no chat template of its own
and borrows the instruct checkpoint's, and the two declare the same 128 added
tokens.

\paragraph{Results}
Table~\ref{tab:baseinstruct} reports the three pairs over the same 200 cases for
both checkpoints of each pair, with paired shifts tested by the Wilcoxon
signed-rank test. On every pair, instruction tuning moves the gap towards the
reserved marker, while the cost of the extra tokens does not shift. No base
checkpoint prefers the reserved marker: Qwen3-1.7B-Base is indifferent, and the
Qwen3-8B and Seed-OSS-36B bases disfavour it. The instruction-tuned side ranges
from $+0.33$ logits on Seed-OSS-36B to $+7.82$ on Qwen3-1.7B, so instruction
tuning moves every pair in the same direction by different amounts.

\begin{table}[h]
\caption{Identity gap and extra-token cost in logits for base and
instruction-tuned checkpoints, 200 cases.}
\label{tab:baseinstruct}
\centering\small
\setlength{\tabcolsep}{7pt}
\begin{tabular}{l l r r K l}
\toprule
\tabhead \textbf{Model} & \textbf{Quantity} & \textbf{Instruct} & \textbf{Base} & \textbf{Shift} & \textbf{Wilcoxon $\boldsymbol{p}$} \\
\midrule
Qwen3-1.7B   & identity gap     & $+7.82$ & $+0.15$ & $\mathbf{+7.67}$ & $3.2\times10^{-25}$ \\
             & extra-token cost & $+0.45$ & $+0.08$ & \ns{$+0.37$} & $0.11$ \\
\midrule
Qwen3-8B     & identity gap     & $+0.68$ & $-0.46$ & $\mathbf{+1.14}$ & $4.6\times10^{-7}$ \\
             & extra-token cost & $+0.21$ & $+0.11$ & \ns{$+0.10$} & $0.20$ \\
\midrule
Seed-OSS-36B & identity gap     & $+0.33$ & $-2.60$ & $\mathbf{+2.92}$ & $3.4\times10^{-29}$ \\
             & extra-token cost & $+0.11$ & $+0.01$ & \ns{$+0.10$} & $0.15$ \\
\bottomrule
\end{tabular}
\end{table}

\subsection{Reasoning Suppression}
\label{app:samebatch}

\paragraph{Both Settings Together}
Reasoning is suppressed through \texttt{enable\_thinking}, a per-prompt argument
of Qwen3's chat template, so the reasoning-on and reasoning-off versions of
\Res, \Spl{} and \Mat{} can run together on the
same cases (Qwen3-8B, 400 cases, five runs, 1536-token budget).
Table~\ref{tab:nothink} gives the result. The difference between the two gaps,
paired within case, is $+41.0$ pp on direct harm (interval $[+34.0, +48.0]$) and
$+25.8$ pp on data stealing (interval $[+19.3, +32.6]$), close to the $+41.7$
and $+25.5$ pp estimated from separate runs. On both attack types the reserved conditions stay level or rise under
suppression while \Spl{} falls by 21 to 39 pp. Suppression also removes
truncation, and the fall of \Spl{} appears as a rise in generations without any
tool call, from $25\%$ to $64\%$ on direct harm. Figure~\ref{fig:mech}b uses the
separate runs, which also include \Pla{} and \Per.

\begin{table}[h]
\caption{Reasoning suppression on Qwen3-8B with both settings run together:
success rates (\%) and identity gaps (pp).}
\label{tab:nothink}
\centering\small
\setlength{\tabcolsep}{7pt}
\begin{tabular}{l l r r}
\toprule
\tabhead \textbf{Attack} & \textbf{Condition} & \textbf{Reasoning on} & \textbf{Reasoning off} \\
\midrule
DH & \Res & 84.3 & 86.8 \\
   & \Mat & 83.5 & 85.5 \\
   & \Spl & 74.7 & 35.7 \\
   & \cellcolor{TabKey}identity gap $\Delta$ & \cellcolor{TabKey}$\mathbf{+8.8}$ & \cellcolor{TabKey}$\mathbf{+49.8}$ \\
\midrule
DS & \Res & 90.8 & 96.2 \\
   & \Mat & 89.9 & 94.9 \\
   & \Spl & 89.5 & 68.7 \\
   & \cellcolor{TabKey}identity gap $\Delta$ & \cellcolor{TabKey}\ns{$+0.4$} & \cellcolor{TabKey}$\mathbf{+26.2}$ \\
\bottomrule
\end{tabular}
\end{table}

\paragraph{Qwen3-32B}
The same comparison at 32B, on 400 direct-harm cases with three runs, gives gaps
of $+17.6$ pp with reasoning and $+34.7$ pp without it, a difference of
$+17.1$ pp against $+41.0$ at 8B. The reserved conditions are again flat while
\Spl{} falls by 15.6 pp. The pattern carries across scale in direction, with a
smaller size because the 32B model's \Spl{} rate stays higher without reasoning
($56.6\%$ against $35.8\%$).

\paragraph{GLM-4.5}
\label{app:nothinkglm}
On GLM-4.5 we applied the same manipulation with three runs per attack type, at
a 384-token budget held fixed across the manipulation. Suppressing reasoning
costs the non-reserved conditions $60\%$ to $99\%$ of their success and the
reserved ones $5\%$ to $16\%$. The gap itself moves by only 2 to 4 pp, because
GLM-4.5's \Spl{} rate is already $12\%$ with reasoning on, which caps how far the
gap can widen.

\section{AgentDojo}
\label{app:agentdojo}

\paragraph{Protocol}
\label{app:adepisode}
We use AgentDojo v1 with its default attack, which writes the payload into every
injection placeholder that the environment offers, and apply the condition's
encoder to every tool-result span in the conversation. A success requires
AgentDojo's own security check to find the injection task's goal state reached
after the tools have run with the arguments the model gave them. Both held-out
splits were fixed before any run. Runs use a 6144-token budget, three repeats
and a $10\%$ floor on the \Res{} rate, fixed in advance. The floor is checked
once per configuration and
repeat, on the \Res{} rate pooled over the suites in that run. It is not a
per-suite exclusion rule, so suites whose own \Res{} rate lies below $10\%$, such
as workspace and Seed-OSS-36B travel, are still reported. The construction
checks of App.~\ref{app:invariants} hold on every tool-result span. Pairs on
which \MatB{} misses \Spl's token count or the marker
position (33 of 409 on Qwen3) are excluded from its contrasts in advance.
Intervals are case-clustered bootstrap intervals over pairs; clustering on the
user task instead gives the same conclusions.

\paragraph{Episode Composition}
The median episode carries 6 untrusted spans on Qwen3-8B and 3 on Llama-3.1.
Llama-3.1's chat template raises an error on any assistant message with more
than one tool call, so we cap each turn at one call in every condition.
$6.6\%$ of Qwen3-8B's episodes reach the generation limit, against none of
Llama-3.1's. On slack and travel \Mat{} truncates more often than \Spl, which
lowers \Mat{} and makes the reported gaps conservative.

\paragraph{Four-Suite Split}
Table~\ref{tab:adpos} reports the held-out split of 409 pairs over all four
suites. Pooled over suites, all three gaps are positive with every lower bound
above zero on both models: $+10.3$ pp $[+7.5, +13.1]$ on Qwen3-8B and $+6.8$ pp
$[+4.7, +9.1]$ on Qwen3-32B for $\Delta$, and $+6.2$ to $+9.5$ pp for the two
position-matched gaps. The cost of the extra tokens is equivalent to zero. Travel
and slack carry the largest gaps on both models, and workspace, where \Res{}
succeeds on under $10\%$ of pairs, carries the smallest on Qwen3-32B.

\begin{table}[h]
\caption{AgentDojo, four-suite held-out split: gap between each matched
condition and \Spl{} (pp).}
\label{tab:adpos}
\centering\small
\setlength{\tabcolsep}{7pt}
\begin{tabular}{l l r K r r}
\toprule
\tabhead \textbf{Model} & \textbf{Suite} & \textbf{Pairs} & \Mat & \MatB & \MatA \\
\midrule
Qwen3-8B  & all four  & 409 & $\mathbf{+10.3}$ & $+9.5$ & $+8.2$ \\
          & banking   & 89  & \ns{$+2.2$} & \ns{$+2.7$} & \ns{$+9.0$} \\
          & slack     & 50  & \ns{$+14.0$} & \ns{$+7.4$} & \ns{$+5.3$} \\
          & travel    & 85  & $\mathbf{+27.4}$ & $+25.6$ & $+14.5$ \\
          & workspace & 185 & $\mathbf{+5.2}$  & $+6.6$  & $+5.8$ \\
\midrule
Qwen3-32B & all four  & 409 & $\mathbf{+6.8}$  & $+6.2$  & $+6.4$ \\
          & banking   & 89  & \ns{$+6.0$} & \ns{$+5.0$} & \ns{$+5.2$} \\
          & slack     & 50  & $\mathbf{+21.3}$ & \ns{$+18.5$} & \ns{$+12.7$} \\
          & travel    & 85  & \ns{$+12.9$} & $+15.1$ & $+15.3$ \\
          & workspace & 185 & \ns{$+0.5$} & \ns{$+0.7$} & \ns{$+1.3$} \\
\bottomrule
\end{tabular}
\end{table}

\paragraph{Three-Suite Split}
A second held-out split of 224 pairs over banking, slack and travel adds
Llama-3.1 and Seed-OSS-36B (Table~\ref{tab:adcells}). \Res{} succeeds on
$2.4\%$ to $41.3\%$ of pairs, a range that includes the $32.05\%$ that
\citet{chatinject} report for the template attack. The gap is positive for all
ten combinations of model and suite and significant in every run for six of
them, including the two primary tests fixed in advance, Qwen3-8B travel and
Llama-3.1 banking. Every suite shows an established gap on at least two models.
The cost of the extra tokens reaches significance in 1 of 30 combinations of
model, suite and run, about what chance gives. Llama-3.1 is evaluated on banking
only, its primary test, a restriction fixed before any held-out run: its low
task completion on travel and slack bounds how often an injection can succeed
there. On Llama-3.1 banking, utility under attack, the fraction of
episodes in which the user's own task is completed, is $26.6\%$, $24.0\%$ and
$21.3\%$ under \Res, \Spl{} and \Mat: attack success falls from $22.5\%$ to
$6.7\%$ under \Spl{} while task completion stays in the same range.

\begin{table}[h]
\caption{AgentDojo, three-suite held-out split: success rates (\%) and identity
gaps (pp). Stars mark the two primary tests fixed in advance.}
\label{tab:adcells}
\centering\small
\setlength{\tabcolsep}{7pt}
\begin{tabular}{l l r r r r K}
\toprule
\tabhead \textbf{Model} & \textbf{Suite} & \textbf{Pairs} & \Res & \Mat & \Spl & $\boldsymbol{\Delta}$ \\
\midrule
Qwen3-8B     & banking            & 89 & 21.0 & 23.6 & 18.4 & \ns{$+5.2$} \\
             & slack              & 50 & 41.3 & 42.0 & 33.3 & \ns{$+8.7$} \\
             & travel$^{\star}$   & 85 & 26.3 & 30.2 & 4.7  & $\mathbf{+25.5}$ \\
\midrule
Llama-3.1-8B & banking$^{\star}$  & 89 & 22.5 & 19.1 & 6.7  & $\mathbf{+12.4}$ \\
\midrule
Qwen3-32B    & banking            & 89 & 17.6 & 21.7 & 13.5 & \ns{$+8.2$} \\
             & slack              & 50 & 24.7 & 26.7 & 6.7  & $\mathbf{+20.0}$ \\
             & travel             & 85 & 21.2 & 17.3 & 1.6  & $\mathbf{+15.7}$ \\
\midrule
Seed-OSS-36B & banking            & 89 & 28.1 & 22.1 & 7.5  & $\mathbf{+14.6}$ \\
             & slack              & 50 & 14.7 & 18.0 & 2.7  & $\mathbf{+15.3}$ \\
             & travel             & 85 & 2.4  & 3.9  & 0.4  & \ns{$+3.5$} \\
\bottomrule
\end{tabular}
\end{table}

\paragraph{Benign Utility}
\label{app:utility}
The defence is the \Spl{} encoding, so its cost on benign traffic is measured by
the same pipeline without an injection task: AgentDojo's 57 banking, slack and
travel user tasks, run with untrusted spans encoded normally and defensively,
three repeats, scored by task utility. The defended condition reaches $69.0\%$
and the undefended one $64.9\%$. The per-repeat differences are $+7.0$, $-3.5$
and $+8.8$ pp, none of them significant, and the design detects a difference of
about 13.6 pp with $80\%$ power.

\section{Coverage Gap and Adaptive Attacks}
\label{app:coverage}

\subsection{Tokenizer Census}
\label{app:census}

\paragraph{The Mitigation}
Hugging Face tokenizers expose
\texttt{split\_\allowbreak special\_\allowbreak tokens=True}, which encodes the
strings of special tokens as ordinary text, and \texttt{tiktoken} exposes
\texttt{disallowed\_\allowbreak special}. Both act on tokens declared \texttt{special:true}
and leave added tokens declared \texttt{special:false} atomic. On Qwen3-8B the
latter include \verb+<tool_call>+, \verb+</tool_call>+, \verb+<tool_response>+
and \verb+</tool_response>+, through which agent frameworks pass tool calls and
untrusted tool output, and the reasoning delimiters \verb+<think>+ and
\verb+</think>+.

\paragraph{Selection}
We walk the Hub's text-generation models in order of downloads and keep every
repository whose tokenizer loads without custom code and exposes a chat
template, with no filter on vendor, family, size or licence, until 400 are kept.

\paragraph{Classification}
For each checkpoint we load the tokenizer and encode the string of every added
token with and without \texttt{split\_special\_tokens=True}, which we call the
flag. A token is left intact by the flag when both encodings agree and the
string remains a single id. Such tokens are then classified by protocol role
rather than by literal string, since families spell the same role differently:
\verb+<tool_call>+ in Qwen3, \verb+<seed:tool_call>+ in Seed-OSS, and full-width
bars in DeepSeek. The tool-protocol class covers tool-call and tool-response
markers and reasoning delimiters such as \verb+<think>+.
Checkpoints whose tokenizers behave identically are grouped
together, which gives 67 distinct configurations.

\paragraph{Reachability}
A token counts only if an attacker can place its id. For each of a user turn, a
system turn, a trailing assistant turn and a generation prompt, the rendered
prompt must contain the id when the token's string is inside the message and
must not contain it otherwise. This excludes tokens that the template emits
anyway. An attacker who
writes such a token inside a tool result still adds a boundary that the template
would not have placed there, so the count is a lower bound.

\begin{figure}[h]
\centering
\includegraphics[width=0.78\linewidth]{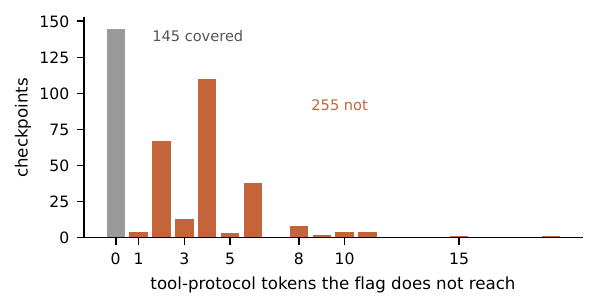}
\caption{Tool-protocol tokens outside the reach of
\texttt{split\_special\_tokens}, per checkpoint, among the 400 most-downloaded
chat models on the Hugging Face Hub. Gray marks checkpoints that the flag fully
covers.}
\label{fig:census}
\end{figure}

\paragraph{Result}
Of the 67 configurations, 33, carrying 255 of the 400 checkpoints ($64\%$),
declare between 1 and 19 reachable tool-protocol tokens that the flag leaves
intact, with a median of 4 (Figure~\ref{fig:census}). Four checkpoints also
leave a reachable role marker intact.
Table~\ref{tab:census} lists the configurations of the families this paper
studies and their close relatives. Seven of these nine configurations, covering
15 of 19 checkpoints, leave 5 to 13 tool-protocol tokens outside the flag's
reach. Llama-3.1, Llama-3.3 and gpt-oss declare no such tokens, which is why the
Llama configurations have no tool-channel variant in App.~\ref{app:toolchan}.
GLM-4.6 ships GLM-4.5's configuration unchanged.

\begin{table}[h]
\caption{Added tokens that \texttt{split\_special\_tokens} leaves intact, and
the subset in the tool-protocol class, which includes reasoning delimiters, for
the families studied here and their close relatives.}
\label{tab:census}
\centering\small
\setlength{\tabcolsep}{7pt}
\begin{tabular}{l r r K}
\toprule
\tabhead \textbf{Checkpoints} & \textbf{Count} & \textbf{Left intact} & \textbf{Tool-protocol} \\
\midrule
DeepSeek-V3.1                         & 1 & 18 & 13 \\
GLM-4.5, GLM-4.6                      & 2 & 14 & 10 \\
Kimi-K2-Thinking                      & 1 & 7  & 7 \\
Qwen3 (8B, 14B, 30B-A3B, 235B-A22B, \ldots) & 7 & 12 & 6 \\
Qwen3.5-4B, Qwen3.5-9B                & 2 & 12 & 6 \\
Seed-OSS-36B-Instruct                 & 1 & 6  & 6 \\
Kimi-K2-Instruct                      & 1 & 5  & 5 \\
Llama-3.1-8B, Llama-3.3-70B           & 2 & 0  & 0 \\
gpt-oss-20b, gpt-oss-120b             & 2 & 0  & 0 \\
\bottomrule
\end{tabular}
\end{table}

\subsection{The Tool Channel}
\label{app:toolchan}
We repeat \Res, \Spl{} and \Mat{} on a forged block built only from a
configuration's \texttt{special:false} tool-protocol tokens, which closes the
current tool response and opens a spoofed second one. The original forged
system turn runs alongside it on the same cases (Table~\ref{tab:toolchan}). The
gap is
present on the uncovered channel for every model that declares such tokens, and
on Qwen3-8B it is larger there than on the covered channel. The cost of the extra
tokens is null on Qwen3-8B and GLM-4.5 and reaches significance in one of three
runs on the other two. The spoofed tool block is a weaker attack than the forged
system turn on every model (last column). On Qwen3-8B
the tool-channel gap was also measured on two further case draws, with $+9.7$
and $+10.3$ pp, significant in every run of each.

\begin{table}[h]
\caption{Identity gap (pp) for a forged block of tool-protocol tokens, which the
standard mitigation leaves intact, and for the original forged system turn,
whose tokens it covers, direct harm, one experiment per model. The last column
is the change in \Res{} success when the tool block replaces the system block.}
\label{tab:toolchan}
\centering\small
\setlength{\tabcolsep}{8pt}
\begin{tabular}{l K r r}
\toprule
\tabhead \textbf{Model} & \textbf{Tool channel} & \textbf{System channel} & \textbf{\Res{}: tool minus system} \\
\midrule
Qwen3-8B     & $\mathbf{+11.7}$ & $+8.4$  & $-25.2$ \\
Qwen3-32B    & $\mathbf{+19.9}$ & $+17.7$ & $-18.0$ \\
GLM-4.5      & $\mathbf{+15.7}$ & $+60.2$ & $-50.7$ \\
Seed-OSS-36B & $\mathbf{+9.4}$  & $+58.8$ & $-45.2$ \\
\bottomrule
\end{tabular}
\end{table}

\subsection{Encoder Behaviour on Benign Traffic}
\label{app:noop}
The source-aware encoder of App.~\ref{app:encoder} encodes untrusted spans with
the tokenizer's special-token matcher removed and nothing else changed. On an
untrusted span that contains no control-token string it emits exactly the same
token ids as the standard encoder, so the defence leaves such spans unchanged.
On real tool traffic, without any model, the defence changes no token in 545 AgentDojo
tool outputs from the banking, slack and travel suites with injections disabled,
or in 17 InjecAgent tool responses, on Qwen3-8B, Llama-3.1 and GLM-4.5 alike.
Text that quotes control markers verbatim, such as
documentation about chat templates, is re-segmented by design.

\subsection{Adaptive Lookalike Search}
\label{app:adaptive}

\paragraph{Candidates}
A lookalike spelling carries no reserved id and the defence does not touch it,
so it remains open to the attacker; MetaBreak \citep{metabreak} searches over
such spellings. For each family we generated candidate replacements for its
forged markers in nine categories: letter substitutions; leetspeak; deletions;
insertions; transpositions; case changes; homoglyphs; whitespace, zero-width and
bracket variants; and, following MetaBreak, ordinary tokens near the reserved
token in input-embedding space. Every candidate was checked on the rendered
prompts of all cases: under the defended encoding it carries no reserved id, it
changes only the marker strings, and the rest of the payload is byte-identical.
Together with six fixed rules written before the search (the first two letters
to \texttt{zz}, which is \Look; the last two letters to \texttt{zz}; the first
letter to its successor; vowels to \texttt{o}; the first two letters swapped;
and the first letter upper-cased, which is \Recase), this gives 115 to 133
candidates per configuration.

\paragraph{Selection and Test}
Every candidate ran once on calibration cases separate from the held-out draw: the
110 direct-harm cases outside the draw of Table~\ref{tab:main}, and 144
data-stealing cases. The best candidate on calibration then ran on the held-out
draw together with \Res, \Spl, \Pla{} and the six fixed rules, with three
repeats. With and without the defence, the attacker's best option is the most
successful spelling available in that setting, and the value of the defence is
the difference between the two best options.

\begin{table}[!htb]
\caption{Best attacker success (\%) with and without the defence, and the
success the defence removes (pp). A neighbour of rank $k$ replaces each marker
with the $k$-th nearest ordinary token to the reserved one in input-embedding
space. Undefended is the attacker's best option without the defence. For the
six-rule columns, the undefended baseline is the best of \Res{} and the six
rules, which differs from the Undefended column only for Qwen3-8B DS
($90.4\%$).}
\label{tab:lksearch}
\centering\footnotesize
\setlength{\tabcolsep}{4.5pt}
\begin{tabular}{l l l r r r r K}
\toprule
\tabhead & & & & \multicolumn{2}{c}{\textbf{Defended best}} & \multicolumn{2}{c}{\textbf{Removed}} \\
\tabhead \textbf{Model} & \textbf{Attack} & \textbf{Best spelling} & \textbf{Undefended} & \textbf{Six rules} & \textbf{Search} & \textbf{Six rules} & \textbf{Search} \\
\midrule
Qwen3-8B     & DH & closing \texttt{|} removed & 85.7 & 74.9 & 81.2 & $+10.8$ & $+4.5$ \\
Llama-3.1-8B & DH & neighbour, rank 14  & 98.0 & 52.8 & 92.2 & $+45.2$ & $+5.8$ \\
GLM-4.5      & DH & neighbour, rank 9   & 72.0 & 26.3 & 70.4 & $+45.7$ & $+1.6$ \\
Seed-OSS-36B & DH & neighbour, rank 8   & 84.8 & 33.0 & 72.6 & $+51.8$ & $+12.2$ \\
Qwen3-8B     & DS & closing \texttt{|} removed & 93.1 & 89.2 & 93.1 & $+1.2$  & $0.0$ \\
\bottomrule
\end{tabular}
\end{table}

\paragraph{Results}
Table~\ref{tab:lksearch} extends Table~\ref{tab:adaptive} of the main text with
data stealing and with the success the defence removes, and shows that the
search recovers most of what the defence removes. Against the six fixed rules the defence removes 1.2 to 51.8 pp, while
against the best searched spelling it removes 0.0 to 12.2 pp. On Llama-3.1 the
best spelling replaces each marker with the fourteenth-nearest ordinary token to
the reserved one and reaches $92.2\%$ on held-out cases, whereas every
character-level variant of the marker stays at $50\%$ to $59\%$ on calibration.
The spelling that works drops the look of the template altogether, consistent
with the negative surface term on Llama-3.1 in App.~\ref{app:armn}. On Qwen3-8B
it removes the closing \texttt{|} of each marker. The selected spellings hold
their calibration rates on held-out cases, and using the five best candidates
instead of one changes the best defended option by at most 2.1 pp.